\documentclass[prx,twocolumn, superscriptaddress, svgnames]{revtex4-2}
\usepackage[T1]{fontenc} 
\usepackage{tikz,tikzscale, relsize, pgfplots, xcolor, amssymb, amsmath}
\usetikzlibrary{backgrounds, arrows, calc, decorations, positioning, decorations.pathmorphing,decorations.pathreplacing, 	decorations.markings, fixedpointarithmetic, fit, patterns, decorations.text, backgrounds, matrix, plotmarks, spy, 3d, shapes.multipart, angles, quotes}
\usepgfplotslibrary{fillbetween}
\pgfplotsset{compat=1.17}
\usetikzlibrary{external}
\usepackage{graphicx}
\usepackage{ragged2e}

\usepackage{array}

\usepackage{arrayjobx}
\graphicspath{{./figsfin/}}
\DeclareGraphicsExtensions{.pdf,.png,.jpg}

\usetikzlibrary{arrows.meta}

\begin{document}
\title{Determination of Magnetic Symmetries by Convergent Beam Electron Diffraction}

\author{O. Zaiets}
\email{o.zaiets@ifw-dresden.de}
\affiliation{Leibniz Institute for Solid State and Materials Research Dresden, Helmholtzstraße 20, 01069 Dresden, Germany}
\affiliation{Institute of Solid State and Materials Physics, TU Dresden, Haeckelstraße 3, 01069 Dresden, Germany}
\author{C. Timm}
\affiliation{Institute of Theoretical Physics, TU Dresden, 01062 Dresden, Germany}
\affiliation{W\"urzburg--Dresden Cluster of Excellence ct.qmat, TU Dresden, 01062 Dresden, Germany}
\author{J. Rusz}
\affiliation{Department of Physics and Astronomy, Uppsala University, 75120 Uppsala, Sweden}
\author{J.-\'{A}. Castellanos-Reyes}
\affiliation{Department of Physics and Astronomy, Uppsala University, 75120 Uppsala, Sweden}
\author{S. Subakti}
\affiliation{Leibniz Institute for Solid State and Materials Research Dresden, Helmholtzstraße 20, 01069 Dresden, Germany}
\author{A. Lubk}
\email{a.lubk@ifw-dresden.de}
\affiliation{Leibniz Institute for Solid State and Materials Research Dresden, Helmholtzstraße 20, 01069 Dresden, Germany}
\affiliation{Institute of Solid State and Materials Physics, TU Dresden, Haeckelstraße 3, 01069 Dresden, Germany}
\affiliation{W\"urzburg--Dresden Cluster of Excellence ct.qmat, TU Dresden, 01062 Dresden, Germany}

\begin{abstract}
Convergent-beam electron diffraction (CBED) is a well-established probe for spatial symmetries of crystalline samples, mainly exploiting the well-defined mapping between the diffraction groups (symmetry group of CBED patterns) and the point-group symmetries of the crystalline sample. In this work, we extend CBED to determine magnetic point groups. We construct all magnetic CBED groups, of which there exist 125. Then, we provide the complete mapping of the 122 magnetic point groups to corresponding magnetic CBED groups for all crystal orientations. In order to verify the group-theoretical considerations, we conduct electron-scattering simulations on antiferromagnetic crystals and provide guidelines for the experimental realization. Based on its feasibility using existing technology, as well as on its accuracy, high spatial resolution, and small required sample size, magnetic CBED promises to become a valuable alternative method for magnetic structure determination.
\end{abstract}

\maketitle
\section{Introduction}

Symmetry has always played an important role in the fundamental understanding and prediction of physical phenomena. In particular, Neumann's principle for solid-state physics connects structural point-group or space-group symmetry of a crystal with the symmetry of its physical properties, such as dielectric, piezoelectric, and elastic linear response tensors. By extension, this principle also holds for magnetic properties, which raises the need for experimental methods for determining symmetries of crystals incorporating the magnetic structure. We introduce a novel comprehensive method for probing magnetic symmetries in magnetically ordered crystalline matter (i.e., ferro-, ferri-, and antiferromagnets) by leveraging electron-diffraction techniques employing a Transmission Electron Microscope (TEM). Its simplicity, spatial resolution down to ten nanometers, and cost-effectiveness are suited to drastically expand the scope of magnetic-symmetry determination currently carried out by neutron diffraction.


Magnetic point and space groups~\cite{Zamorzaev1953}, also referred to as Shubnikov groups, provide a fundamental classification underlying all magnetic or spin-related properties of crystalline solids~\cite{Rodriguez-Carvajal2019}, such as ferro-, ferri- and antiferromagnetism~\cite{Blundell2001}, altermagnetism~\cite{Simejkal2022}, topological insulation~\cite{Tokura2019}, and semimetallicity~\cite{Armitage2018,Watanabe2018}. Magnetic space groups extend the crystallographic space groups by combining spatial symmetries with time-reversal symmetry, which connects opposite spin orientations, and provide compatibility conditions for magnetic properties following Neumann's principle. Therefore, experimental probes of magnetic point-group and space-group symmetries, analogous to probes of crystallographic symmetries, are indispensable for studies of solid-state magnetism, with neutron diffraction being the main probe~\cite{Izyumov1991}. A direct determination of the magnetic symmetry from the symmetries of a neutron diffraction pattern is, however, hampered by Friedel's law that holds for scattering  within the Born approximation: Opposite reflections in a neutron diffraction pattern are always symmetric even for noncentrosymmetric space groups. Motivated by this and other limits of neutron diffraction, such as limited spatial resolution and relatively large required crystal sizes, we establish an alternative technique based on electron diffraction in this paper. We introduce a method for the determination of magnetic point groups that allows direct measurement of symmetries with a spatial resolution reaching several nanometers, referred to as ``magnetic convergent-beam electron diffraction.''

\begin{figure}[h]
    \centering{
   \includegraphics{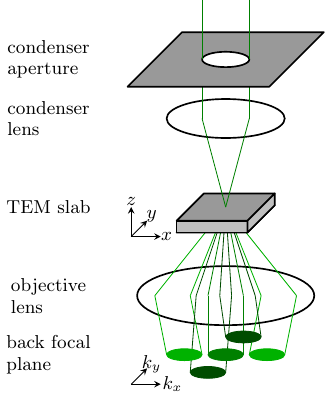}}
    \caption{Minimal optical model of a CBED setup. The beam converges in the sample plane, while the beam-limiting aperture is in the front focal plane of the condenser lens. The CBED pattern is recorded in the back focal plane (far field), where diffracted beams appear as disks because of the convergent illumination. Sample space as well as focal plane (far field) coordinates are indicated.}
    \label{fig:CBED rays}
\end{figure}

Convergent-beam electron diffraction (CBED) is a transmission-electron-microscopy (TEM) technique that focuses, i.e., converges, an electron beam, typically of $60$--$300\,\mathrm{kV}$ acceleration voltage, to spot sizes of the order of $10$ nanometers on a thin TEM sample slab and records the transmitted diffraction pattern of Bragg disks~\cite{Kossel1939, Morniroli2012, Jacob2012}, see Fig.\ \ref{fig:CBED rays}. Such CBED patterns have their own specific diffraction symmetries, i.e., mirror or rotational symmetries, which depend on structural symmetry groups and the zone-axis orientation of the TEM sample slab~\cite{Buxton1976, Tanaka2010}. More specifically, the CBED patterns directly inherit the point-group symmetries of the scattering potential of the TEM sample slab.  Additional Friedel pair symmetries are not present because electron scattering on crystalline samples typically falls outside of the Born approximation regime~\cite{Goodman1968}.
In other words, the beam electrons scatter multiple times---also referred to as dynamical scattering---instead of just once, as described by the first Born approximation and also referred to as kinematical scattering. Opposite reflections then loose their symmetry in noncentrosymmetric scattering potentials because of the nonsymmetric intermediate scattering phases of the diffracted beams. 
The point-group symmetries of the TEM sample slab are given by the intersection of the original bulk crystal symmetries with the geometric symmetries of the slab. In other words, all crystal symmetries changing the normal direction of the slab, referred to as $z$ in the following, are broken with the notable exception of $z$reversal, i.e., mirror symmetry with respect to $xy$-plane at the center of the slab, and combinations of $z$reversal with in-plane point group symmetries. Therefore, the CBED symmetry of a slab of particular orientation corresponds to the in-plane crystallographic symmetries of the slab plus possible $z$-reversal symmetries, including combinations of in-plane and $z$-reversal symmetries. This stipulates a well-defined mapping between CBED pattern symmetries and structural point-group symmetries~\cite{Buxton1976}, which may be uniquely inverted provided that a sufficient number of CBED patterns of slabs in various zone-axis orientations is recorded. Numerous studies have employed this inverse mapping to study symmetries of crystalline matter down to very small length scales~\cite{Francesconi1998, Tsuda1999, Enidjer2009, Tsuda2013a, Shi2013, Hayashida2020}, revealing the presence of strain, ferroelectric polarization, or chirality. An extension of this mapping to space-group symmetries requires considerable effort regarding the incompatibility of slab symmetries with translational symmetries in the $z$ direction. Although we will touch upon this topic below, our main focus in this work is on point-group symmetries.

While the conventional CBED technique focuses solely on structural symmetries of the crystal imprinted on the electrostatic scattering potential, a growing number of electron-scattering studies highlight the impact of the magnetic field. In an early study, Shen and Laughlin~\cite{Shen1990} showed the breaking of structural CBED symmetries by the presence of the strong uniaxial macroscopic magnetic field in ferromagnetic PrCo$_5$. Later, Loudon~\cite{Loudon2012} revealed the existence of purely magnetic electron-diffraction spots in parallel-beam electron diffraction on antiferromagnetic NiO, where the antiferromagnetic unit cell is doubled compared to the structural one. This work paved the way for the detection of antiferromagnetic order in the Transmission Electron Microscope (TEM). Recently, Kohno \textit{et al.}\ \cite{Kohno2022} demonstrated the sensitivity of atomic-resolution differential phase contrast (another convergent-beam electron-diffraction technique, where first moments of diffraction patterns are analyzed) to the antiferromagnetic structure of Fe$_2$O$_3$. These and similar observations~\cite{Krizek2022, Nguyen2023, Tanigaki2024} suggest the experimental possibility to detect the generally weak electron-diffraction effects due to atomic magnetic fields~\cite{Edstrom2016a} in antiferromagnets, which possess no macroscopic magnetic field. The latter is the main impetus for the following generalization of the conventional CBED symmetry analysis by taking into account symmetries of the magnetic crystal structure. 

We tackle the problem by first reviewing diffraction of electrons on electric and magnetic potentials in thin crystalline slabs (i.e., crystalline electron-transparent TEM samples), establishing the relation between symmetries of the slab and those of the scattered electrons. In Sec. \ref{sec:CBEDgroups}, we then provide a classification of all possible CBED diffraction symmetry groups by employing group theory. In the following Sec. \ref{sec:CBEDG_MPG_map} the complete map between CBED diffraction groups and magnetic point groups of the sample is provided. The corresponding tables for all crystal classes are shown in Appendix \ref{app:mPGvsmDG}. In Sec.  \ref{sec:examples} exemplary computational case studies for two different antiferromagnets are discussed. Finally, Sec. \ref{sec:experiment} elaborates on the experimental feasibility of magnetic CBED.

\section{Electron diffraction at electric and magnetic potentials}
\label{sec:diffraction}

Electron beams accelerated by voltages in the $60\,\mathrm{kV}$ to $300\,\mathrm{kV}$ range attain relativistic velocities (e.g., $v\approx 0.8\, c$ at $300\,\mathrm{kV}$ acceleration voltage) and are predominantly scattered by very small angles (less than $100\,\mathrm{mrad}$) relative to the optical axis in a transmission electron microscope. 
In this small-angle, or paraxial scattering, limit, the effects of spin through spin-orbit coupling and Zeeman coupling are negligible, meaning the spinor nature of the electron wave function can be ignored. As a result, spin does not influence the diffraction, allowing us to treat the electron's wave function as a scalar.
%
Moreover, the resulting scalar wave function $\psi$ may be decomposed into a fast carrier wave function and an envelope wave function $\Psi$ that is slowly varying along the optical axis, i.e., $\psi(\boldsymbol{r}_{\bot},z) = e^{ik_{0}z}\, \Psi(\boldsymbol{r}_{\bot},z)$ with $\boldsymbol{r}_{\bot}=\left(x,y\right)$ denoting the in-plane slab coordinates and $k_0$ the wave number of the fast carrier wave. Here, the time-dependent oscillation has already been separated, i.e., we only consider eigensolutions with a certain electron energy $E$. Inserting the above decomposition into the Klein-Gordon equation for scalar relativistic particles and neglecting a couple of terms that are small due to the high beam energy, the evolution of the envelope wave function along the optical axis can be described by the paraxial Schr\"odinger equation~\cite{Rother2009, Edstrom2016} (see Appendix~\ref{app:parax} for a detailed derivation)
\begin{align}
i\,\frac{\partial\Psi}{\partial z}
  &= \left(\frac{1}{2k_{0}}\left(i\boldsymbol{\nabla}_{\bot}
    + \frac{e}{\hbar}\, \boldsymbol{A}_{\bot}\right)^2
    + \frac{e}{\hbar}\, A_{z}
    - \frac{e}{\hbar v}\, \Phi\right) \Psi 
  \nonumber \\
&\equiv \hat{H}_\bot \Psi , 
\label{eq:hhplNeelGaus}
\end{align}
where $e$ is the elementary charge, $\boldsymbol{A}=(\boldsymbol{A}_\bot,A_z)$ the vector potential, and $\Phi$ the scalar potential. Spatial arguments of potentials and envelope wave function have been omitted in order to simplify the notation. The vector potential is fixed by the Coulomb gauge in the following considerations and generally leads to effects several orders of magnitude smaller than the diffraction due to the scalar potential. As a consequence of the large wave vector of the fast-beam electrons, one may further simplify the above equation by neglecting in-plane vector-potential terms containing $\boldsymbol{A}_{\bot}$. 

The symmetries of the solution of the above equation in the absence of the vector potential, $\boldsymbol{A} = 0$, have been exploited for structural symmetry determination using CBED based on the following argument. A CBED pattern corresponds to the absolute square of the electron wave function in the far field behind the object (coordinates $\boldsymbol{k}_\perp=\left(k_x,k_y\right)^\mathrm{T}$). Therefore, the CBED discs appear at integer multiples of unit vectors of the crystal's reciprocal lattice according to Bragg's law, see Fig.\ \ref{fig:CBED rays}. The electron wave function, on the other hand, inherits symmetries from the paraxial Hamiltonian $\hat{H}_\bot$ and hence symmetries of the electric scattering potential $\Phi$, which leads to a direct relationship between symmetries of the latter and those of the CBED pattern~\cite{Bird1989}. Given the similar mathematical structure of the above equation with and without magnetic vector potential, this reasoning may be also adopted to derive the relationship between CBED pattern symmetries and magnetic point-group symmetries. Because of the additional sign change of the magnetic vector potential upon time-reversal, the latter has to be included in the symmetry considerations, when considering CBED on magnetic materials. 

The structure of Eq.\ (\ref{eq:hhplNeelGaus}) corresponds to a two-di\-men\-sio\-nal time-dependent Schr\"odinger equation, i.e., a $\left(2+1\right)$-dimensional theory, with $z$ and the paraxial Hamiltonian $\hat{H}_{\bot}$ taking the place of the time coordinate and the usual Hamiltonian, respectively~\cite{Berry1971, Gratias1983}. In the following, we will take advantage of this analogy, allowing for a substantial portion of the time-dependent Schr\"odinger equation's properties and solutions to be transferred to paraxial scattering.

The full solution of the paraxial equation (\ref{eq:hhplNeelGaus}) for a given initial wave at the entrance plane $z_0$ of the slab can be formally obtained as 
\begin{equation}
\Psi(z) = \underbrace{Z
  \exp\left(-i\int_{z_{0}}^{z} dz'\,
  \hat{H}_{\bot}\left(z'\right)
  \right)}_{\equiv \hat{K}(z,z_{0})} \Psi(z_{0}) ,
\end{equation}
where $Z$ denotes the $z$-ordering directive in analogy to Dyson's time-ordering directive and $\hat{K}(z,z_{0})$ is the $z$-evolution operator (or propagator) corresponding to the time-evolution operator. The intensity $I(\boldsymbol{k}_{\bot})$ recorded in a CBED pattern is obtained by applying the propagator to the wave function, i.e., the convergent probe beam, in reciprocal space,
\begin{equation}
    I\left(\boldsymbol{k}_{\bot}\right)
    = \left|\iint d^2k'_\bot\,
    \hat{K}(\boldsymbol{k}_{\bot},z,\boldsymbol{k}'_{\bot},z_{0})\,
    \Psi(\boldsymbol{k}'_{\bot},z_0)\right|^2 .
\end{equation}
In full analogy to time evolution, the $z$ evolution is unitary, conserving the norm of the wave function along $z$, because the Hamiltonian $H_{\bot}$ is hermitian. 
Moreover, any in-plane symmetries (rotations, mirror reflections, and translations in the $xy$ plane) leave the propagator invariant because the corresponding unitary symmetry transformations $\hat{T}_{\bot}$ commute with the Hamiltonian $\hat{H}_{\bot}$ and hence with the propagator, $[\hat{T}_{\bot},\hat{K}] = 0$. Translational symmetries correspond to phase factors in reciprocal space, which are not directly visible as diffraction symmetries but instead lead to absences of diffraction peaks. Hence, only in-plane rotation and mirror symmetries, i.e., the in-plane point-group symmetry of the slab, translate to symmetries of the CBED pattern,
provided that the initial probe wave function also has these symmetries. Rotational and mirror symmetries are present for a circularly  symmetric converged electron beam used in CBED experiments. To match these beam symmetries with the point group symmetries of the sample (i.e., to align the beam propagation axis with the high-symmetry crystallographic axis of the sample), the beam direction has to be carefully adjusted in the experiment. Here, the beam position (center of mass of the beam) within a unit cell, i.e., whether it strictly coincides with a high-symmetry axis or not, is typically irrelevant if the size of the convergent illumination spans several unit cells.

Another notable analogy between the two-dimensional time-dependent and the paraxial $z$-dependent Schr\"o\-din\-ger equations concerns the reversal of the propagation parameter. Time reversal corresponds to an antiunitary operator consisting of complex conjugation of the scalar wave function. The same complex conjugation is applied to the envelope wave function under $z$ reversal, which can be shown along the same lines as for the time-dependent Schr\"odinger equation.
Moreover, if the paraxial Hamiltonian $\hat{H}_{\bot}$ is $z$-reversal invariant, the propagator and hence the CBED pattern will remain the same after $z$ reversal of the slab if the following important condition is fulfilled: The upper and lower face of the slab need to be identical crystal planes. If this condition is not satisfied the crystal slab breaks $z$-reversal symmetry even if the bulk crystal has this symmetry.

In addition, the presence of nonzero magnetization and vector potential gives rise to another symmetry operation---magnetization or time reversal---inherited by the CBED pattern. The full symmetries, including time reversal, of the CBED patterns of a magnetic sample may be obtained by the same logic than those of nonmagnetic ones. We start from the magnetic space group symmetries of the crystal. The slab geometry breaks spatial symmetries involving $z$, possibly except for $z$-reversal, and imprints these symmetries onto the CBED pattern recorded along the normal axis of the slab. Exploiting this principle, magnetic CBED in a Transmission Electron Microscopoe can be a viable alternative to neutron diffraction for the determination of magnetic point groups of magnetic samples provided that the mapping of magnetic CBED symmetry groups (also referred to as magnetic diffraction groups (DG)) to magnetic point groups can be worked out and the presence of magnetic CBED symmetries can be determined experimentally. 

In the following, we construct the complete magnetic CBED group (magnetic DG) from group-theoretical principles and provide the complete mapping between magnetic CBED groups and magnetic point groups. We then conduct scattering simulations involving numerical solutions of Eq.\ (\ref{eq:hhplNeelGaus}) for antiferromagnetic samples in order to verify the theory. Subsequently, we discuss the experimental challenges for implementing magnetic CBED.

\section{Magnetic CBED groups} \label{sec:CBEDgroups}

In this section, we discuss the possible point groups of magnetic CBED patterns. Details and lists of all groups are relegated to Appendix~\ref{app:grouptheory}. The construction relies on the following main assumptions: (a) Time reversal and \textit{z} reversal both act as antiunitary transformations, as explained in Sec.\ \ref{sec:diffraction}. (b) Both square to the identity since fermionic properties of electrons can be disregarded for relativistic beams. This point has also been discussed in Sec.\ \ref{sec:diffraction}. In other words, we do not require double groups. (c) Time reversal and \textit{z} reversal commute with all other structural point-group transformations. (d) The two also commute with each other.

Starting from the magnetic point group of the bulk crystal, we first need to construct the subgroup of elements that leave the slab geometry invariant, as discussed in Sec.\ \ref{sec:diffraction}.
This leads to the $10$ two-dimensional structural point groups, in Schoenflies notation, $C_n$ and $D_n$ with $n=1,2,3,4,6$. It is easy to check explicitly that all elements of these groups commute with \textit{z} reversal. The fact that \textit{z} reversal is antiunitary does not affect the commutation relations \footnote{One can find a purely real representation of all group elements and thereby show that the additional complex conjugation introduced by making \textit{z} reversal antiunitary does not change the commutation relations.}. Time reversal commutes with all possible structural transformations in any case~\cite{Wigner1931}. These arguments justify assumptions (c) and~(d).

Including the antiunitary time-reversal operation $\mathcal{T}$ leads to $31$ two-dimensional magnetic point groups in total~\cite{Litvin2013}: First, the $10$ groups, $G$, noted above, which do not contain time reversal and are called ``monochromatic.'' Second, $10$ groups $G + G\mathcal{T}$, which result from adding time reversal $\mathcal{T}$ as a generator. These groups are called ``gray.'' Third, $11$ groups that are created by identifying a halving subgroup $H$ of a monochromatic group $G$ and forming the group $H + (G \setminus H)\mathcal{T}$, i.e., by multiplying the coset $G \setminus H$ by time reversal. These groups are called ``dichromatic.'' For an introduction to magnetic point groups see, for example,~\cite{Bradley_1968, DDJ2008}.

We now denote the antiunitary reversal of the \textit{z} coordinate by $\mathcal{Z}$. This operation is antiunitary, squares to the identity, and commutes with all other structural symmetry operations. Hence, $\mathcal{Z}$ and $\mathcal{T}$ have the same algebraic properties and the theory of point groups including $\mathcal{Z}$ is analogous to the theory of point groups including $\mathcal{T}$. Thus by including $\mathcal{Z}$ but not $\mathcal{T}$, we obtain $10$ ``$\mathcal{Z}$ gray'' groups and $11$ ``$\mathcal{Z}$ dichromatic'' groups.

Novel possibilities arise if $\mathcal{T}$ and $\mathcal{Z}$ are both present, either by themselves or only in combination with structural transformations or with each other. The latter combination, which we denote by
\begin{equation}
\Omega \equiv \mathcal{T} \mathcal{Z}  = \mathcal{Z} \mathcal{T} ,
\end{equation}
is \emph{unitary}. The properties of $\mathcal{T}$ and $\mathcal{Z}$ imply that $\Omega$ squares to the identity and commutes with all possible group elements.

For any of the $10$ structural groups $G$, we can now construct CBED groups by applying one or more of the following steps: 

(1) We can add one of the operations $\mathcal{T}$, $\mathcal{Z}$, or $\Omega$ as a new generator. We can also add two of them as new generators. Since the product of any two operations is the third one, this only produces $10$ additional groups $G + G\mathcal{T} + G\mathcal{Z} + G\Omega$. These groups are both $\mathcal{T}$ gray and $\mathcal{Z}$ gray.

(2) If $G$ has a halving subgroup $H$, then its coset $G\setminus H$ can be multiplied by one of the operations $\mathcal{T}$, $\mathcal{Z}$, or $\Omega$. 

(3) If $G$ has two distinct halving subgroups $H_1$ and $H_2$, then the corresponding cosets $G\setminus H_1$ and $G\setminus H_2$ can be multiplied by two distinct operations out of $\{\mathcal{T}, \mathcal{Z}, \Omega\}$.

As detailed in Appendix~\ref{app:grouptheory}, we end up with $125$ magnetic CBED groups in total. Table \ref{table:G.summary} summarizes the results. We conclude this section with two remarks: First, two groups can be distinct even if they have the same structure, i.e., correspond to the same formal group. For example, the groups $C_2$ and $D_1$ correspond to the same formal group but are distinct because the symmetry transformation involved is physically different---a rotation and a mirror reflection, respectively.
Second, the orders of the groups are easily read off from the second column, noting that $H$ is a halving subgroup of $G$ and $Q$ is a quartering subgroup of $G$.

\begin{table*}
    \caption{\label{table:G.summary}Summary of CBED groups. The construction, the resulting type of the groups, and their number is shown. $G \in \{C_n, D_n|n=1,2,3,4,6\}$ is a structural point group, $H$ is a halving subgroup of $G$, and $Q$ is a quartering subgroup of $G$, i.e., a halving subgroup of a halving subgroup. Moreover, $\mathcal{T}$ denotes time reversal, $\mathcal{Z}$ denotes \textit{z} reversal, and $\Omega = \mathcal{TZ}$ denotes their product. $c_n$ denotes the highest-order rotation in $G$ and $m_x$ the mirror reflection $x\to -x$.}
    \begin{ruledtabular}
    \begin{tabular}{cccr}
    \multicolumn{2}{c}{Construction} & Type & Number \\ \hline
    structural & $G \in \{C_n, D_n|n=1,2,3,4,6\}$ & monochromatic & $10$ \\
    add $\mathcal{T}$ & $G + G\mathcal{T}$ & $\mathcal{T}$ gray & $10$ \\
    add $\mathcal{Z}$ & $G + G\mathcal{Z}$ & $\mathcal{Z}$ gray & $10$ \\
    add $\Omega$ & $G + G\Omega$ & monochromatic & $10$ \\
    add all three & $G + G\mathcal{T} + G\mathcal{Z} + G\Omega$ &
        $\mathcal{T}$, $\mathcal{Z}$ gray & $10$ \\
    $\mathcal{T}$ dichromatic & $H + (G\setminus H)\mathcal{T}$ & $\mathcal{T}$ dichromatic & $11$ \\
    $\mathcal{Z}$ dichromatic & $H + (G\setminus H)\mathcal{Z}$ & $\mathcal{Z}$ dichromatic & $11$ \\
    $\Omega$ pseudo-dichromatic & $H + (G\setminus H)\Omega$ & monochromatic & $11$ \\
    (pseudo-)dichromatic, add $\mathcal{T}$ &
        $H + (G\setminus H)\mathcal{Z} + [H + (G\setminus H)\mathcal{Z}]\mathcal{T}$ &
        $\mathcal{T}$ gray & $11$ \\
    (pseudo-)dichromatic, add $\mathcal{Z}$ &
        $H + (G\setminus H)\mathcal{T} + [H + (G\setminus H)\mathcal{T}]\mathcal{Z}$ &
        $\mathcal{Z}$ gray & $11$ \\
    dichromatic, add $\Omega$ &
        $H + (G\setminus H)\mathcal{T} + [H + (G\setminus H)\mathcal{T}]\Omega$ &
        $\mathcal{T}$, $\mathcal{Z}$ gray & $11$ \\
    double dichromatic with $c_n\mathcal{T}$ &
        $Q + Q c_n \mathcal{T} + Q m_x \Omega + Q c_n m_x \mathcal{Z}$ &
        $\mathcal{T}$, $\mathcal{Z}$ dichromatic & $3$ \\
    double dichromatic with $c_n\mathcal{Z}$ &
        $Q + Q c_n \mathcal{Z} + Q m_x \Omega + Q c_n m_x \mathcal{T}$ &
        $\mathcal{T}$, $\mathcal{Z}$ dichromatic & $3$ \\
    double dichromatic with $c_n\Omega$ &
        $Q + Q c_n \Omega + Q m_x \mathcal{T} + Q c_n m_x \mathcal{Z}$ &
        $\mathcal{T}$, $\mathcal{Z}$ dichromatic & $3$ \\ \hline
    &&& $125$
    \end{tabular}
    \end{ruledtabular}
\end{table*}

\section{Relation between magnetic CBED groups and magnetic point groups}\label{sec:CBEDG_MPG_map}

The group-theoretical considerations in Sec.\ \ref{sec:CBEDgroups} have provided a list of all 125 magnetic CBED groups possible by combining two-dimensional structural point-group symmetry operations with time reversal (and thus magnetization reversal) and $z$ reversal. The missing piece required for determining magnetic point groups from CBED measurements is a mapping from the 122 three-dimensional magnetic point groups of magnetic materials to the 125 magnetic CBED groups, which we will provide in this section, with all mappings tabulated in Appendix~\ref{app:mPGvsmDG}.

\begin{figure}
\centering
\centering{\includegraphics{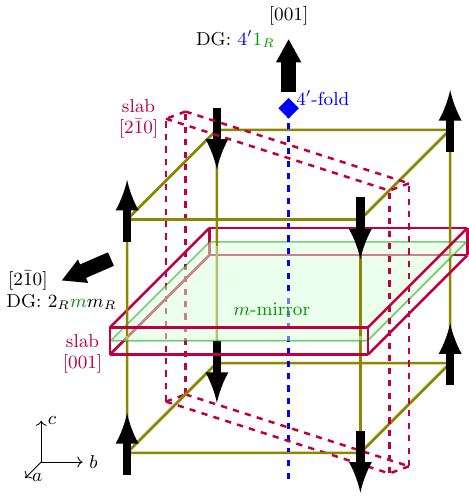}}
\caption{Magnetic point group $4'/m$ symmetry broken by the slab geometry of a thin TEM sample of tetragonal symmetry. "`DG'" stands for diffraction group, i.e., magnetic CBED group.}
\label{fig:4povermmDGvsmPG}
\end{figure}

When the electron beam is scattered by the thin TEM sample, the magnetic point-group symmetry is broken by the TEM slab geometry as discussed in Sec.\ \ref{sec:diffraction}. An example for the case of magnetic point group $4'/m$ is presented in Fig.~\ref{fig:4povermmDGvsmPG}. Here, the prime denotes time reversal as usual. This group is generated by the fourfold rotation about the principal axis multiplied by time reversal, $4'$, and the reflection in the mirror plane perpendicular to that axis, $m$.

We first consider a slab with surface normal oriented along the $[001]$ crystallographic axis, i.e., the fourfold symmetry axis. Thus, this slab orientation is compatible with the point group $4'/m$. The mirror reflection in the horizontal plane is then mapped to the $z$-reversal symmetry, which we denote by $1_R$, following the conventions from CBED literature~\cite{Buxton1976}. The corresponding magnetic CBED group is thus $4'1_R$.

On the other hand, a slab with normal along $[2\bar{1}0]$ is incompatible with all operations that involve rotations by $90^\circ$. This reduces the symmetry of the slab to the point group $2/m$ (see Fig.~\ref{fig:4povermmDGvsmPG}). Describing this point group within the coordinate system defined by the slab / CBED experiment (i.e., $z$-axis parallel to beam slab normal / beam direction), the mirror symmetry along $c$ becomes a mirror along $y$ and the twofold rotation around $c$ becomes a $z$-reversal combined with a $x$-mirror. Hence, the twofold rotation is mapped to $m_R$. Noting that the product of $m$ and $m_R$ is a two-fold rotation within the plane multiplied by $z$-reversal, i.e. $2_R$, the magnetic CBED group of the $[2\bar{1}0]$ slab turns out to be $2_Rmm_R$.

These examples illustrate the general procedure we apply to generate the complete mapping of all magnetic point groups: First, we  generate a representative magnetic structure for a certain magnetic point group. In the second step, we analyze the preserved symmetry operations for a certain slab orientation. Finally, we map the reversal of the coordinate normal to the slab to the antiunitary $z$-reversal. To cover all possibilities, i.e., all symmetry classes of slab orientations, we probe all classes of orientations that yield different conventional CBED symmetries, as tabulated in the literature~\cite{Buxton1976}. It turns out that all possible 125 magnetic CBED groups from Sec.\ \ref{sec:CBEDgroups} are generated by applying the above algorithm for all 122 three-dimensional magnetic point groups. The above procedure was carried out with the help of a software package provided by De Graef~\cite{DeGraef2010}.

In Appendix~\ref{app:mPGvsmDG}, we present the full mapping for all distinct beam directions, categorized along crystal classes. This mapping may be used to determine which CBED data, e.g., which slab orientations need to be prepared and recorded using CBED, are sufficient to unambiguously reconstruct the magnetic-point-group symmetries of the studied sample, similarly to the well-known procedure for the determination of the structural point group by conventional CBED. In the following section, we discuss exemplary electron scattering simulations corroborating the group-theoretical analysis.

\section{Computational examples}\label{sec:examples}

High-resolution TEM data as well as electron-diffraction data from magnetic crystal structures can be reliably and quantitatively reproduced by numerical integration of Eq.\ (\ref{eq:hhplNeelGaus}) using a split-operator algorithm, referred to as ``multislice'' in the TEM literature, as described in~\cite{Edstrom2016}. Here, the electrostatic scattering potentials were assembled from parameterized atomic potentials described in~\cite{Peng1996}, including experimental values of the Debye-Waller factors~\cite{lee_magnetoelastic_2016,emery_variable_2011} and the related complex absorptive part of the potential employing a temperature of $300\,\mathrm{K}$ and an acceleration voltage of $300\,\mathrm{kV}$ \cite{Castellanos-Reyes2023}. The parametrized magnetic vector potentials reported in~\cite{lyon_parameterization_2021} were used. In order to reduce computation time, we did not simulate the thermal diffuse background due to inelastic phonon or magnon scattering in electron diffraction because that background does not affect the symmetry of elastic CBED patterns. The background is important, however, when considering whether weak magnetic diffraction effects can be reliably measured utilizing a certain electron dose in the experiment \cite{Snarski-Adamski_simulations_2023}. In all simulations, the starting wave function (i.e., the convergent TEM probe beam) has been normalized in intensity to unity when summed over all sampling points of the diffraction pattern.
For the exemplary simulations we picked two anitferromagnetic crystals that are suited to illustrate the different aspects of the proposed method. 

\subsection{LaMnAsO}

\begin{figure}
\centering{\includegraphics{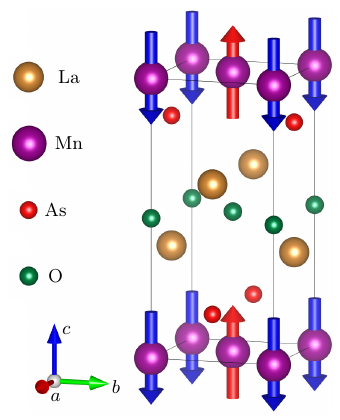}}
\caption{Crystal structure of LaMnAsO with directions of the Mn magnetic moments.}
\label{fig:LMAOspins2}
\end{figure}

\begin{figure*}
\centering{\includegraphics{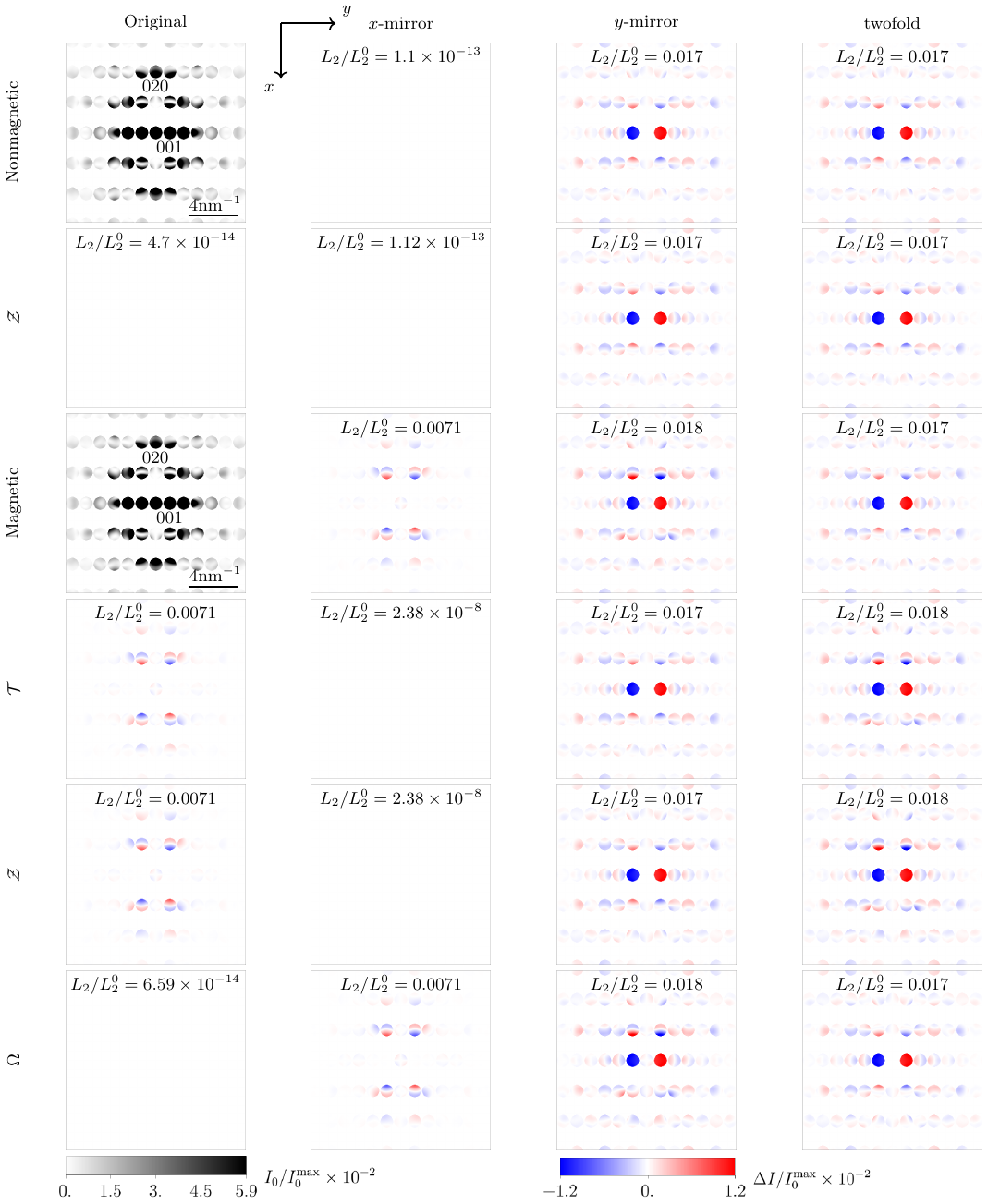}}
\caption{CBED patterns pertaining to a LaMnAsO slab (thickness $\approx 100\,\mathrm{nm}$) in $\left[100\right]$ orientation. Reference nonmagnetic and magnetic CBED patterns are shown at beginning of first and third row. Otherwise, differences of CBED patterns are depicted. Rows indicate (1) nonmagnetic, (2) $z$-reversed nonmagnetic, (3) magnetic, (4) time- (magnetization-) reversed magnetic, (5) $z$-reversed magnetic, (6) time- and $z$-reversed magnetic structure.
Columns indicate (1) as computed, (2) $x-$mirrored, (3) $y-$mirrored, and (4) $180^\circ$-rotated CBED patterns, see text.}
\label{fig:LMAO SF}
\end{figure*}

The first computational example is LaMnAsO, which is an antiferromagnet with magnetic space group $P4'/n'm'm$ and lattice parameters $a=b=4.114\,\mbox{\AA}$ and $c=9.0304\,\mbox{\AA}$~\cite{McGuire2016}. The magnetic moments of Mn atoms are oriented along the easy axis $c$ and are arranged in the antiferromagnetic pattern shown in Fig.\ \ref{fig:LMAOspins2}~\cite{McGuire2016}. The symmetries of LaMnAsO that are probed by magnetic CBED are described by the point group $4'm'm$. We furthermore note that the point group $4'm'm$ is equivalent to $4'mm'$ obtained by rotating the unit cell through $45^{\circ}$ about $[001]$. The latter conventions have been used for the magnetic CBED groups and their mappings to magnetic point groups in Table \ref{table:Tetragonal2}. 

Studying Table \ref{table:Tetragonal2}, zone axis orientation $[110]$, corresponding to $[100]$ for the unit-cell convention used in the simulation, is indeed sufficient to uniquely distinguish between the magnetic point-group symmetries compatible with the structural tetragonal $4mm$ symmetry. We have therefore conducted CBED scattering simulations on slabs oriented along the $[100]$ zones axis. To avoid overlap of the CBED disks at an acceleration voltage of $300\,\mathrm{kV}$, we have chosen a convergence semiangle (i.e., semiangle subtended by converged electron beam) of $1\,\mathrm{mrad}$ that is smaller than half the distance between reciprocal lattice points. In order to observe a sufficient level of detail within the Bragg disks, we needed to construct supercells of large lateral size. In the case of $[100]$ zone axis, we have built a supercell of dimensions $31c \times 68a$, which is approximately $28\,\mathrm{nm} \times 28\,\mathrm{nm}$, with a grid of $5580 \times 5576$ pixels. The maximal sample thickness was set to $243a$, which is approximately $100\,\mathrm{nm}$.

In total, six multislice calculations have been carried out. Two were nonmagnetic (i.e., magnetic vector potential neglected), one for the original structure model (first row in Fig.\ \ref{fig:LMAO SF}) and the other for a $z$-reversed structure model (second row) to corroborate the well-known structural symmetry map, which is also included in Table \ref{table:Tetragonal2} as structural point groups. Furthermore, four magnetic calculations have been performed for: (third row) the original structure model (fourth row) with reversed magnetic moments (i.e., time-reversed situation), (fifth row) the $z$-reversed structure with (sixth row) reversed magnetic moments. Here, it is worth mentioning that magnetic-moment components that are perpendicular (parallel) to the $z$-axis need to be reversed (stay constant) under $z$-reversal because the magnetic moments are axial vectors.

The simulated CBED patterns pertaining to the nonmagnetic and the magnetic CBED simulations are displayed in the first and third panel of the left column of Fig.\ \ref{fig:LMAO SF}, respectively. Their intensity is displayed as gray scale. The difference between the nonmagnetic and magnetic simulation is visually not noticeable as it amounts to relative intensity variations of the order of $10^{-2}$ to $10^{-3}$, corroborating the weak magnitude of magnetic scattering effects. The CBED patterns exhibit the typical inner structure of the Bragg disks imprinted by dynamical (multiple) scattering. We note, however, that due to the limited sampling of directions in the simulations as well as the saturation of inner Bragg disks in the chosen color scale not all fine details are visible.

Whether a particular symmetry is present in the computed CBED patterns was then checked by computing the difference $\Delta I=I-I_0$ between the CBED pattern obtained by applying the corresponding symmetry transformation and the original one. These difference patterns are displayed in Fig.\ \ref{fig:LMAO SF} using a red-blue color scale. Additionally we also provide the Euclidean $L_2$ norm of $\Delta I$ normalized by Euclidean norm of $I_0$. Inspecting the nonmagnetic results in the first and second row of Fig.\ \ref{fig:LMAO SF}, we see that the structural CBED symmetries contain $x$-mirror reflection and $z$ reversal, denoted by $m1_R$, which is predicted by Table \ref{table:Tetragonal2} for the point group $4mm$. Other symmetry operations possible in tetragonal systems, such as a twofold rotation, produce a difference in the percent range and can hence be ruled out.

Turning to the magnetic simulations (third to sixth rows), we observe $m'1'_R$ magnetic CBED symmetry, which again agrees with our predictions, see Table \ref{table:Tetragonal2}. The violation of both $m$ and $1_R$ symmetries, present in the structural simulation, is only slightly weaker compared to structurally absent symmetries of the third and fourth column. The magnitude of symmetry preservation or violation due to magnetic scattering effects therefore should be accessible in the experiment. Note, however, that signal-to-noise ratios (SNR) and hence CBED acquisition times must be chosen to large enough to overcome the noise background of structural Bragg disks, as discussed in Sec.~\ref{sec:experiment}. A possible reduction of the SNR issue is naturally provided by antiferromagnetic materials that possess a magnetic unit cell that is larger (typically by a factor of 2) than the structural one. This class of antiferromagnets exhibit purely magnetic Bragg reflections without a structural contribution and thus a significantly reduced background in these discs. In the next subsection, we will discuss a prominent member of this family, antiferromagnetic NiO.

\subsection{NiO}\label{app:NiO}

\begin{figure}
\centering{\includegraphics{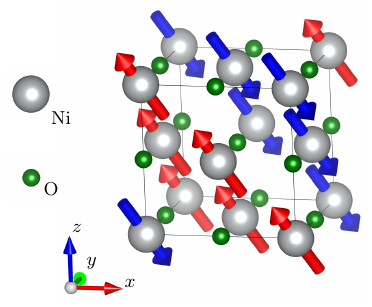}}
\caption{Crystal structure of NiO with directions of the magnetic moments. Note that only one structural cubic unit cell is displayed. The cubic antiferromagnetic unit cell is twice as large in each direction. The magnetic monoclinic unit cell, on the other hand, lies askew within the cubic cell and has the same volume.}
\label{fig:NiOspins}
\end{figure}

NiO is a Mott-insulating antiferromagnet below a N\'eel temperature of $523\,\mathrm{K}$~\cite{Roth1958}. Its paramagnetic phase has cubic $Fm\bar{3}m$ symmetry with lattice constant $a=4.17\,\mbox{\AA}$, which becomes monoclinic $C_c2/c$ (BNS magnetic space group notation \cite{belov1957shubnikov}) in the antiferromagnetic phase~\cite{Cairns1933} (Fig.~\ref{fig:NiOspins}). The latter involves a doubling of the magnetic unit cell. The monoclinic axis of the magnetic unit cell points into the $[1\bar{1}0]_c$ direction (the cubic notation is indicated by the subscript \textit{c}) and the spins are oriented along the easy axis $[11\bar{2}]_c$, which is perpendicular to the monoclinic axis. It is also important to note that NiO undergoes a spin-flop transition at an external magnetic field of $1.54\,\mathrm{T}$, where spins reorient perpendicular to the field direction~\cite{Saito_1980}. Indeed, Loudon~\cite{Loudon2012} detected the purely magnetic reflections of the spin-flopped phase by conventional broad-beam electron diffraction, where the objective lens typically imposes external fields well above the spin-flop field. The magnetic symmetry of the spin-flopped phase has not been reported yet. 

We performed two sets of simulations, one for the original phase and one for the spin-flopped phase. The slab orientation for the original phase was the monoclinic axis $[1\bar{1}0]_c$. The slab orientation for the spin-flopped phase was $[11\bar{2}]_c$, corresponding to the monoclininc $a$-axis, and the spin orientation in the spin-flopped phase was along $[1\bar{1}0]_c$ (monoclinic $b$-axis). The slab thickness was set to approximately $102\,\mathrm{nm}$. Again, we conducted four magnetic CBED simulations, covering all possible combinations of $z$ and time reversal. The supercell size in these simulations was set to $289.252\,\mathrm{\mathring{A}}\times 283.406\,\mathrm{\mathring{A}}$ that was sampled with $5760\times5760$ points.

\begin{figure}
\centering{\includegraphics{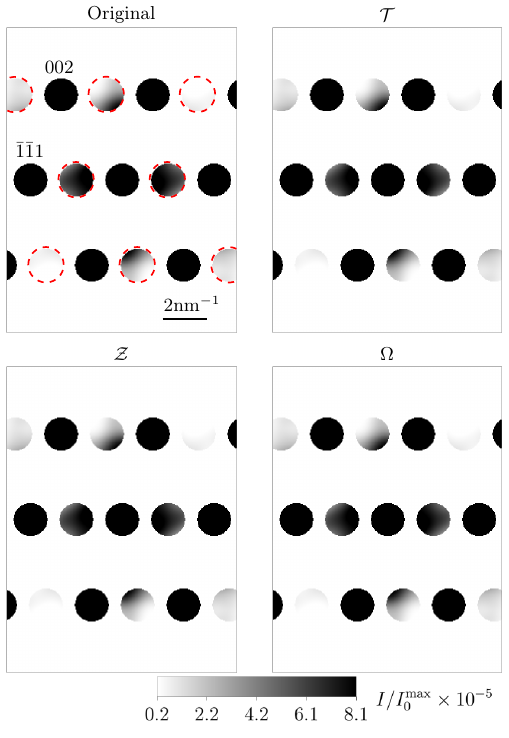}}
\caption{CBED patterns pertaining to the magnetic ground state NiO slab in $\left[1\bar{1}0\right]_c$ orientation. Purely magnetic reflections are indicated by dashed red circles.}
\label{fig:NiO gray}
\end{figure}

\begin{figure}
\centering{\includegraphics{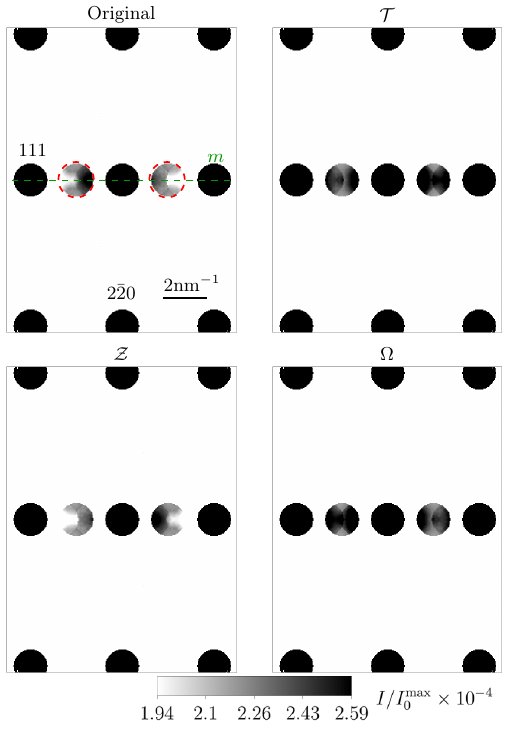}}
\caption{CBED patterns pertaining to the spin-flopped NiO slab in $\left[11\bar{2}\right]_c$ orientation. The spin orientation is $\left[1\bar{1}0\right]_c$. Purely magnetic reflections are indicated by dashed red circles.}
\label{fig:NiO SF}
\end{figure}

In contrast to LaMnAsO, we can now directly read off the symmetries from the purely magnetic reflections displayed in Figs.\ \ref{fig:NiO gray} and \ref{fig:NiO SF}. Here, in order to display their internal structure, a gray scale was chosen that saturates the structural Bragg disks. The CBED pattern of the original phase exhibits twofold rotational symmetry as well as both $z$-reversal and time-reversal symmetries, i.e., $21_R1'$ in total, see Fig.\ \ref{fig:NiO gray}. Inspecting Table \ref{table:Monoclinic}, this symmetry corresponds to the gray point group $2/m1'$, which corresponds to the magnetic point group symmetries obtained by stripping all translations from the NiO space group $C_c2/c$. Indeed, the glide plane translation along the monoclinic $c$ direction does not break the symmetry for the slab oriented along monoclinic $b$-axis since $c$ is perpendicular to $b$, and leaves the slab invariant. Consequently, the glide plane symmetry translates into a simple mirror symmetry denoted by $/m$ in the $2/m1'$ diffraction group mentioned above. Similarly, the $1'_c$ symmetry (abbreviated by the subscript $c$ attached to $C$ centering in the space group symbol), i.e., the combination of time reversal and in-plane translation along $c$ corresponds to a simple time reversal symmetry $1'$ since the translation leaves the slab invariant. 

In the spin-flopped case, each CBED pattern exhibits a mirror ($m$) symmetry with respect to the $x$ axis, see Fig.\ \ref{fig:NiO SF}. Moreover, an $m_R$ symmetry is discernible by comparing the $z$-reversed CBED patterns in the upper and lower row of Fig.\ \ref{fig:NiO SF}. Similarly, we can verify the presence of a $2_R$ rotational symmetry. Further symmetries including those involving time reversal are absent. Consequently, the CBED symmetry observed in this orientation is $2_Rmm_R$, which can be associated to the monoclinic magnetic point group $2/m$. Indeed, inspecting the magnetic unit cell, we observe that the space group symmetry of the spin-flopped phase is $C_c2/m$, which contains the point group transformations $2/m$. The combined translation and time-reversal symmetry $1'_c$ (again abbreviated by the $c$ index in the space group notation) is broken in the $a$-oriented slab because the $a$ and $c$ axes are not perpendicular, i.e., the $c$ translation is not in plane.

\section{Experimental Magnetic CBED}\label{sec:experiment}

Having established the theoretical framework for the determination of magnetic point-group symmetries from CBED patterns, we elaborate on possible and existing experimental realizations. We identify eight main challenges and discuss their impact as well as mitigation strategies:
\begin{enumerate}
  \item Signal-to-noise ratio: The previous considerations show that the modifications of CBED patterns due to magnetic scattering are small for antiferromagnets (ferromagnets will be discussed further below). Local intensity modulations in the CBED discs of LaMnAsO at $300\,\mathrm{kV}$ acceleration voltage and approximately $100\,\mathrm{nm}$ slab thickness that indicate the presence or absence of magnetic-point-group symmetries are of the order of 0.1-1\% with respect to the maximal intensity in the CBED pattern. To measure such a weak modulation above the noise level, one should thus employ low-noise detectors of high dynamic range (i.e., direct electron detectors \cite{MacLaren2020}) with a detection quantum efficiency close to unity, effectively reducing the noise to the inevitable Poissonian noise of the detection process. For the example of LaMnAsO, a beam current of $1\,\mathrm{nA}$ (that is common for CBED) and an acquisition time of $0.4\,\mathrm{s}$ yield a maximal magnetic modulation within the Bragg discs that is more than five times larger than the standard deviation of the detection (shot) noise, which is deemed sufficient for detection according to the Rose criterion~\cite{Rose1974} (lower SNR requirements are also employed frequently though).  
  If studying antiferromagnets with a doubled magnetic unit cell that give rise to purely magnetic Bragg reflections, one may furthermore focus on these reflections because they do not have a strong nonmagnetic background. However, for the example of the spin-flopped NiO, a current of $1\,\mathrm{nA}$  in combination with an acquisition time of $32\,\mathrm{s}$ is necessary to exceed a signal-to-noise level of five for measuring a modulation within the two purely magnetic Bragg discs. The larger acquisition time in this case stems from the weaker magnetic diffraction in NiO compared to LaMnAsO and the sizable thermal diffuse background due to inelastic phonon scattering. We finally also note that the relative magnitude of the magnetic signal depends on the thickness of the slab, the acceleration voltage of the electron beam, the zone-axis orientation, and the material under consideration in a complex way due to dynamic scattering, see, e.g., \cite{Snarski-Adamski_simulations_2023}, typically allowing for some optimization of, e.g., the slab thickness, to further increase the relative magnitude of the magnetic CBED disc modulations.  
   
 \item Inelastic electron scattering (excitation of phonons, plasmons, etc.) leads to a background in experimental CBED patterns, effectively lowering the signal-to-noise ratio of the magnetic Bragg signal. A significant part of this background that cannot be removed by filtering inelastically scattered electrons using an energy filter is due to inelastic phonon scattering, referred to as thermal diffuse scattering (TDS). To assess the magnitude of the TDS background, we carried out electron-diffraction simulations including TDS within the framework of the frozen phonon approximation, see Appendix~\ref{app:TDS}. The results show that the intensity of TDS background at room temperature in NiO at the simulated slab thickness is approximately 10 times larger than the magnetic signal and hence does play a crucial role in experimental visibility of magnetic disks and their symmetries. To facilitate the measurement of magnetic symmetries in the purely magnetic Bragg discs with an SNR of 5 above the TDS background intensity, an acquisition time of $32$ seconds employing beam current in the range of nanoamps is required, which is experimentally  feasible. The acquisition time may be further reduced by cooling the specimen, thereby reducing the TDS background. For instance, cooling down from room temperature to 30~K will results in reduction of required exposure times (doses) by approximately  one order of magnitude, which can be seen in the following way: Measuring 10 times shorter leads to 10 times less magnetic signal, but the background goes down up to 100 times---10 times due to shorter acquisition time and additional 10 times due to 10 times reduced temperature, thus maintaining the same SNR as at 300~K at 10 times longer acquisition. Moreover, as TDS goes down, more electrons will scatter elastically instead, further increasing the SNR. Noting, furthermore, that the TDS background in NiO is comparatively large, due to the rather light atoms involved,  the presence of the TDS background may have less impact on the measurement of magnetic symmetries by magnetic CBED in other materials.
 
\item Thickness gradients of the TEM lamella, strain gradients, adsorbates on the surface, as well as the beam damage may complicate the local analysis of symmetries. They typically lead to a rather diffuse diffraction intensity (background), which breaks the symmetry of CBED patterns. It is difficult to assess the impact of such effects as they crucially depend on details such as zone axis, material composition, adsorbate composition, radiation damage susceptibility, etc. A very coarse approximation may be derived by considering the intensity of the background at the position of a CBED disc and the ratio of thickness of the sample to the relative thickness increase due to a gradient, or the thickness of adsorbates layer, or the the thickness of the damaged layer, respectively.  Assuming a typical thickness of the lamella for magnetic CBED of 100 nm and of the adsorbate or damaged layer of 1 nm, and assuming that the background is constantly spread out over the considered CBED regime, similar background levels as the thermal diffuse background at room temperature are obtained, posing another serious challenge for the technique. Suitable countermeasures to reduce these local symmetry breaking effects are averaging the signal over several measurement points, thereby averaging out impact of adsorbates, gradients, or damaged surface layers, careful cleaning of the sample surfaces, cooling to reduce beam damage, and careful sample preparation through focused ion beam milling, to avoid thickness gradients. 

\item The in-plane magnetic-point-group symmetries of the sample translate to CBED symmetries if the circular symmetric electron probe is well aligned with the crystallographic orientation of the slab. The accuracy of this alignment procedure is limited by the precision of the beam tilt coils of the TEM, typically allowing for fine tuning of beam tilt angles down to $100\,\mu\mathrm{rad}$ and slightly below. Thus a residual beam tilt in that range that can break point-group symmetries has to be taken into account when analyzing the presence of magnetic CBED symmetries. To asses the magnitude of such symmetry-breaking effects, we carried out CBED simulations incorporating small beam tilts with respect to crystalline zone axes, see Appendix~\ref{app:Tilt}. The results show that small residual beam tilts result in small shifts of Bragg disks. The artificially induced symmetry breaking may be suppressed by generally excluding a small outer rim of the CBED discs of approximately $100\,\mu\mathrm{rad}$ in the symmetry analysis, as employed for conventional CBED \cite{LeBeau2010,Xu2018,Oberaigner2023}. 
\item For the determination of magnetic symmetries, the experimental realization of magnetization (i.e., time) reversal without changing the magnetic structure is helpful---but not necessary, see below. There are several methods facilitating magnetization reversal, depending on the particular magnetic material to be studied: Firstly, one can briefly heat up the sample above its N\'eel or Curie temperature. If there is no bias for one N\'eel vector orientation this would allow magnetization reversal in a randomized way. Secondly, short application of an external field  may trigger a spin-flop transition in antiferromagnets, after which a random magnetization reversal may occur if no bias is present. Thirdly, one may trigger domain-wall motion by applying currents or external fields \cite{Wadley_electrical_2016, Wadley18, Amin2023} and thereby change the magnetic domain orientation in the observed area.
Note, however, that one may entirely circumvent the experimental realization of magnetization (time) reversal by recording a sufficient number of CBED patterns along different high-symmetry zone axes. For example, in the case of LaMnAsO, the magnetic point group $4'm'm$ may be uniquely determined from CBED patterns recorded in one or maximally two of the high-symmetry zone axes, like $\langle100\rangle$ and $\langle110\rangle$, see Table \ref{table:Tetragonal2}, if the underlying structural point group $4mm$ is known, e.g., from high-temperature electron diffraction or X-ray measurements. 
\item Standard CBED and other diffraction experiments in the TEM are carried out with the sample immersed in the strong magnetic field of the objective lens of up to $2.5\,\mathrm{T}$, which permits very short camera lengths and hence recording of large CBED patterns without significant aberrations. If the influence of the magnetic field from the objective lens alters the magnetic structure under consideration, e.g., because it triggers a spin-flop transition in an antiferromagnet or reorients the magnetization in a ferromagnet, a zero-field adaption of the CBED setup is required. Depending on the TEM (i.e., number and strength of lenses in condenser and projective), this may be achieved by switching off the objective lens and employing the condenser and projective system of the TEM to facilitate required beam convergence and short camera lengths. If the existing optics is insufficient, additional instrumental modifications of the TEM, e.g., a zero-field objective lens or additional preobjective or postobjective lenses are suited to provide the required CBED conditions ~\cite{Shibata2019}.
\item The $z$-reversal symmetries are generally difficult to observe experimentally because the prepared crystal slab is breaking these symmetries when the upper and lower crystal plane are not identical. As a result, one typically records CBED patterns of multiple slabs of noncolinear crystal orientations, rendering the map of diffraction symmetries to point group symmetries uniquely defined without directly probing $z-$reversal symmetries \cite{Buxton1976}.
\item The proposed magnetic CBED can only provide magnetic point-group symmetries. The point-group transformations being part of glide plane and screw axis are typically not visible, except for some special slab orientations, where the corresponding translations are within the slab plane (see above example of spin-flopped NiO). Consequently, the detection of space-group symmetries requires additional care and may involve more sophisticated methods such as inspection of kinematically forbidden reflections or higher-order Laue zones, similar to conventional CBED~\cite{Gjonnes1965}.  
\end{enumerate}

The above list indicates that experimental realization of magnetic CBED will be very challenging. Most likely, artificial symmetry breaking by thickness gradients, adsorbates and beam damage are the biggest obstacles. Notwithstanding, the above arguments also suggest that doses and hence acquisition times required to achieve sufficient signal to noise in antiferromagnets are in reach. First experimental proofs for the feasibility of probing magnetic symmetries by CBED have already been reported for ferromagnets, where magnetic diffraction is more pronounced compared to compensated antiferromagnets. In a pioneering experimental CBED study on the uniaxial ferromagnet PrCo$_5$, Shen and Laughlin \cite{Shen1990} showed how the structural CBED symmetry is reduced by the presence of magnetic scattering. The structural space group of PrCo$_5$ is $P6/mmm$, which would give rise to $6mm$ CBED symmetry along the $\left[001\right]$ and $\mathrm{2mm}$ along the $[110]$ axis, see Table \ref{table:Hexagonal2}. The experimentally observed CBED symmetry along $\left[110\right]$ is, however, clearly reduced to $m$. The theory developed above allows to explain this result. Inspecting the assignment tables for hexagonal magnetic point groups, Tables \ref{table:Hexagonal1} and \ref{table:Hexagonal2}, two ferromagnetic magnetic point groups \cite{Schmid1973} that are compatible with $6/m$ symmetry and ferromagnetic ordering  exhibit $m$ CBED symmetry. The first is $6/m$ (magnetic CBED symmetry $2_Rmm_R$), while the second is $6/mm'm'$ (magnetic CBED symmetry $2'mm'1'_R$). Recording a second CBED pattern along $\left[110\right]$ with magnetization direction reversed would allow to probe the presence of, e.g., the $2'$ symmetry and hence experimental distinction between the $6/m$ and $6/mm'm'$ symmetries. The second experimental example pertains to CBED of La$_{1-x}$Sr$_x$MnO$_3$ by Tsuda \textit{et al.}\ \cite{Tsuda2001}.  The main result is the identification of a triclinic superlattice reconstruction, presumably due to orbital ordering. Additionally, by inspecting multiple CBED patterns along different beam directions, a symmetry of $\bar{1}$ consistent with a CBED symmetry of $2_R$ along all axes, see Table \ref{table:Triclinic}, could be revealed. Indeed, $\bar{1}$ is the only triclinic ferromagnetic magnetic point group \cite{Schmid1973}, necessitating no additional measurement involving magnetization reversal or different slab orientations to uniquely determine the magnetic symmetry in this case.

\section{Summary}

In this work, we have introduced magnetic CBED for the unambiguous determination of magnetic-point-group symmetries in magnetic crystalline materials. We have classified all magnetic point-group symmetries of the diffraction patterns in terms of 125 magnetic CBED groups. The subsequent mapping of all magnetic point groups to corresponding magnetic CBED groups allows the determination of the latter from a finite series of CBED experiments on slabs of different crystal orientation. A successful experimental realization of the magnetic CBED requires overcoming several experimental challenges, including artificial symmetry breaking by beam damage and adsorbates. Our calculation indicate, however, that the signal level (electron dose) required for a successful experimental realization of the proposed method for elucidating the symmetries in antiferromagnets is achievable in modern TEMs. First experimental realizations of this measurement principle have been demonstrated for ferromagnets previously. 

Theoretically, simplicity and cost-effectiveness should make magnetic CBED an intriguing alternative to neutron diffraction for magnetic-symmetry determination. The high spatial sensitivity of magnetic CBED, allowing to probe very small volumes down to about $10\,\mathrm{nm} \times 10\,\mathrm{nm} \times 100\,\mathrm{nm}$, should facilitate the probing of magnetic symmetries for magnetic micro- and nanoparticles, polycrystalline magnetic matter, magnetic thin films, and magnetic van der Waals materials, that cannot be measured with neutron diffraction. Moreover, the high spatial resolution down to about $10\,\mathrm{nm}$ in principle facilitates mapping of symmetries across phase and domain boundaries, or close to crystal defects, by recording magnetic CBED patterns at different beam positions. As a note of caution, we again emphasize that all these considerations hinge particularly on the beam stability of the sample that is required to achieve sufficient signal to noise ratio for probing the magnetic symmetry.

\section{Acknowledgements}

The authors are grateful to Juri Barthel for useful discussions. Financial support by the Deutsche Forschungsgemeinschaft through Collaborative Research Center SFB 1143, projects A04 and C04, project id 247310070, and the W\"urzburg--Dresden Cluster of Excellence ct.qmat, EXC 2147, project id 390858490, is gratefully acknowledged. J. R. and J.-\'{A}. C.-R. acknowledge the support of the Swedish Research Council (grant no.\ 2021-03848), the Olle Engkvist's Foundation (grant no.\ 214-0331), and the Knut and Alice Wallenberg Foundation (grant no.\ 2022.0079). The simulations were enabled by resources provided by the National Academic Infrastructure for Supercomputing in Sweden (NAISS) at the NSC Centre, partially funded by the Swedish Research Council through grant agreement no.\ 2022-06725.

\appendix

\section{Detailed derivation of the paraxial scattering equation} 
\label{app:parax}

Since the effects of the electron spin, such as spin-orbit coupling and Zeeman coupling, are negligible for small-angle scattering of high-energy beam electrons their energy eigenstates are well described by a stationary Klein-Gordon equation, minimally coupled to static electric and magnetic potentials (arguments have been omitted),
\begin{equation}
    \left(E+e\Phi\right)^{2}\psi=\left(c^{2}\left(-i\hbar\nabla+e\mathbf{A}\right)^{2}
    + m_{e}^{2}c^{4}\right)\psi .
\end{equation}
Because of to high energy $E$ of highly accelerated electrons in a Transmission Electron Microscope the electric potential squared term can be neglected, which is referred to as the high-energy approximation~\cite{Fujiwara1961}. After a couple of rearrangements and dividing by $E=\gamma m_{e}c^{2}$, where $\gamma = 1/\sqrt{1-v^2/c^2}$ is the Lorentz factor, this high-energy limit of the Klein-Gordon equation takes a form that is mathematically equivalent to the stationary Schr\"odinger equation~\cite{Rother2009}
\begin{equation}
    \frac{\gamma^{2}-1}{2\gamma}\, m_{e}c^{2}\psi
    = \frac{1}{2\gamma m_{e}}\left(-i\hbar\nabla+e\mathbf{A}\right)^{2}\psi-e\Phi\psi .
\label{eq:KG3}
\end{equation}
As a consequence of the high kinetic energy of the beam electrons, typical targets (samples) brought into their path do not deflect the electrons by large angles. This observation motivates the paraxial approximation, where the wave is separated into a fast plane wave and a slowly varying envelope function in the propagation direction $z$,
\begin{equation}
    \psi(\boldsymbol{r}_{\bot},z) = \Psi(\boldsymbol{r}_{\bot},z)\,
    e^{ik_{0}z} ,
\end{equation}
where $\boldsymbol{r}_{\bot}$ denotes the two-dimensional coordinates in the $xy$ plane and $k_0=\gamma m_ev/\hbar$ the wave number of the fast carrier wave. Upon insertion of this ansatz into Eq.\ (\ref{eq:KG3}), one can neglect the second-order derivative of the slowly varying envelope function with respect to $z$ as well as first-order derivatives with respect to $z$ multiplied with the small $A_{z}$. Both approximations are motivated by the large $k_{0}$. Moreover, $A^2_{z}$ can safely be dropped for the same reason. One finally obtains the paraxial Schr\"odinger equation
\begin{align}
i\,\frac{\partial\Psi}{\partial z}
  &= \left(\frac{1}{2k_{0}}\left(-i\boldsymbol{\nabla}_{\bot}
    + \frac{e}{\hbar}\, \boldsymbol{A}_{\bot}\right)^2
    + \frac{e}{\hbar}\, A_{z}
    - \frac{e}{\hbar v}\, \Phi\right) \Psi .
\end{align}

\section{Details of the construction of magnetic diffraction groups} \label{app:grouptheory}

In this Appendix, we explain the construction of magnetic CBED groups in detail. As noted in Sec.\ \ref{sec:CBEDgroups}, $10$ possible structural point groups exist for a slab. These groups do not contain any antiunitary elements and are called monochromatic. In addition, the magnetic CBED groups can contain the antiunitary time-reversal operation $\mathcal{T}$, the antiunitary \textit{z} reversal $\mathcal{Z}$, or their unitary product $\Omega$, each either as a new generator or multiplying all elements of the coset $G \setminus H$ of a halving subgroup $H$ of the structural group $G$.

Let $G$ denote any of the structural point groups $C_n$ and $D_n$ with $n=1,2,3,4,6$, where we use the standard Schoenflies notation for two-dimensional point groups (rosette groups). Adding $\mathcal{T}$ as a new generator leads to the $10$ groups $G + G\mathcal{T}$, which we now call ``$\mathcal{T}$ gray'' to distinguish them from the following class. Instead adding $\mathcal{Z}$ leads to the $10$ groups $G + G\mathcal{Z}$, which we call ``$\mathcal{Z}$ gray.'' Finally, adding $\Omega$ as a new generator leads to the $10$ groups $G + G\Omega$. These groups only contain unitary elements and are thus monochromatic. We can also add two of $\mathcal{T}$, $\mathcal{Z}$, or $\Omega$ as new generators. Since the product of any two of these is the third, this only produces $10$ additional groups $G + G\mathcal{T} + G\mathcal{Z} + G\Omega$. These groups are both $\mathcal{T}$ gray and $\mathcal{Z}$ gray. The groups constructed so far are listed in Tables \ref{table:G.mono.gray} and~\ref{table:G.Omega}.

\begin{table}
    \caption{\label{table:G.mono.gray}Structural monochromatic (first and second column), $\mathcal{T}$ gray (third column), and $\mathcal{Z}$ gray (fourth column) CBED groups. The Schoenflies notation is given for the monochromatic groups. The Hermann--Mauguin notation~\cite{HM_notation}
    is presented for all groups, where $1'$ refers to the time-reversal operation $\mathcal{T}$ and $1_R$ denotes the \textit{z}-reversal operation $\mathcal{Z}$.}
    \begin{ruledtabular}
    \begin{tabular}{cccc}
    \multicolumn{2}{c}{Monochromatic groups} & $\mathcal{T}$ gray groups &
      $\mathcal{Z}$ gray groups \\ \hline
    $C_1$ & $1$ & $11'$ & $11_R$ \\
    $C_2$ & $2$ & $21'$ & $21_R$ \\
    $C_3$ & $3$ & $31'$ & $31_R$ \\
    $C_4$ & $4$ & $41'$ & $41_R$ \\
    $C_6$ & $6$ & $61'$ & $61_R$ \\
    $D_1$ & $m$ & $m1'$ & $m1_R$ \\
    $D_2$ & $2mm$ & $2mm1'$ & $2mm1_R$ \\
    $D_3$ & $3m$ & $3m1'$ & $3m1_R$ \\
    $D_4$ & $4mm$ & $4mm1'$ & $4mm1_R$ \\
    $D_6$ & $6mm$ & $6mm1'$ & $6mm1_R$
    \end{tabular}
    \end{ruledtabular}
\end{table}

\begin{table}
    \caption{\label{table:G.Omega}Monochromatic CBED groups containing the product $\Omega = \mathcal{T}\mathcal{Z}$ of time and \textit{z} reversal (first column) and CBED groups that are both $\mathcal{T}$ gray and $\mathcal{Z}$ gray (second column). The Hermann--Mauguin notation~\cite{HM_notation}
    is presented for all groups. $1'_R$ denotes the product operation $\Omega$. Note that $1_R'$ is not given explicitly for the $\mathcal{T}$ and $\mathcal{Z}$ gray groups since it is generated by $1'$ and $1_R$.}
    \begin{ruledtabular}
    \begin{tabular}{cc}
    Monochromatic groups with $\Omega$ & $\mathcal{T}$ and $\mathcal{Z}$ gray groups \\ \hline
    $11'_R$ & $11'1_R$ \\
    $21'_R$ & $21'1_R$ \\
    $31'_R$ & $31'1_R$ \\
    $41'_R$ & $41'1_R$ \\
    $61'_R$ & $61'1_R$ \\
    $m1'_R$ & $m1'1_R$ \\
    $2mm1'_R$ & $2mm1'1_R$ \\
    $3m1'_R$ & $3m1'1_R$ \\
    $4mm1'_R$ & $4mm1'1_R$ \\
    $6mm1'_R$ & $6mm1'1_R$
    \end{tabular}
    \end{ruledtabular}
\end{table}

Moreover, for any structural group $G$ that has a halving subgroup $H$, the elements of the coset $G \setminus H$ can be multiplied by $\mathcal{T}$, $\mathcal{Z}$, or $\Omega$. For time reversal $\mathcal{T}$, this leads to the standard dichromatic groups $H + (G \setminus H)\mathcal{T}$, of which there are $11$. These are listed in Table \ref{table:G.dichro}. As shown in this table, the allowed structural groups have zero, one, or two halving subgroups. Performing the same operation with the \textit{z} reversal operation $\mathcal{Z}$, we obtain another $11$ dichromatic groups $H + (G \setminus H)\mathcal{Z}$. Finally, using $\Omega$, we obtain $11$ groups $H + (G \setminus H)\Omega$, which are monochromatic since $\Omega$ is unitary but could be called ``pseudo-dichromatic'' to highlight their structure. All these groups are also included in Table~\ref{table:G.dichro}.

\begin{table}
    \caption{\label{table:G.dichro}Dichromatic CBED groups based on time reversal $\mathcal{T}$ (first and second column). The Schoenflies notation $G(H)$ for the group $H + (G \setminus H)\mathcal{T}$ and the Hermann--Mauguin notation~\cite{HM_notation}
    are given. The third and fourth column show the $\mathcal{Z}$ dichromatic CBED groups and the ``pseudo-dichromatic'' (but actually monochromatic) CBED groups based on $\Omega$. In the Hermann--Mauguin notation, a prime means that the corresponding generator is multiplied by the time-reversal operation $\mathcal{T}$, a subscript ``\textit{R}'' denotes multiplication by $\mathcal{Z}$, and a prime combined with a subscript ``\textit{R}'' denotes multiplication by $\Omega$.}
    \begin{ruledtabular}
    \begin{tabular}{cccc}
    \multicolumn{2}{c}{$\mathcal{T}$ dichromatic} & $\mathcal{Z}$ dichromatic &
      Pseudo-dichromatic \\[-1.5ex]
    \multicolumn{2}{c}{groups} & groups & groups \\ \hline
    $C_2(C_1)$ & $2'$ & $2_R$ & $2'_R$ \\
    $C_4(C_2)$ & $4'$ & $4_R$ & $4'_R$ \\
    $C_6(C_3)$ & $6'$ & $6_R$ & $6'_R$ \\
    $D_1(C_1)$ & $m'$ & $m_R$ & $m'_R$ \\
    $D_2(D_1)$ & $2'mm'$ & $2_Rmm_R$ & $2'_Rmm'_R$ \\
    $D_2(C_2)$ & $2m'm'$ & $2m_Rm_R$ & $2m'_Rm'_R$ \\
    $D_3(C_3)$ & $3m'$ & $3m_R$ & $3m'_R$ \\
    $D_4(D_2)$ & $4'mm'$ & $4_Rmm_R$ & $4'_Rmm'_R$ \\
    $D_4(C_4)$ & $4m'm'$ & $4m_Rm_R$ & $4m'_Rm'_R$ \\
    $D_6(D_3)$ & $6'mm'$ & $6_Rmm_R$ & $6'_Rmm'_R$ \\
    $D_6(C_6)$ & $6m'm'$ & $6m_Rm_R$ & $6m'_Rm'_R$
    \end{tabular}
    \end{ruledtabular}
\end{table}

Interestingly, dichromatic groups with respect to $\mathcal{T}$ are automatically also dichromatic with respect to $\mathcal{Z}$ and vice versa. To see this, consider a $\mathcal{T}$ dichromatic group $M = H + (G \setminus H)\mathcal{T}$. Define another group
\begin{equation}
\tilde G \equiv H + (G\setminus H)\Omega .
\end{equation}
$H$ is a halving subgroup of $\tilde G$. Now we construct the $\mathcal{Z}$ dichromatic group based on $\tilde G$ and $H$:
\begin{align}
\tilde M &= H + (\tilde G\setminus H)\mathcal{Z} = H + (G\setminus H)\Omega\mathcal{Z}
  \nonumber \\
&= H + (G\setminus H)\mathcal{T} = M .
\end{align}
Thus the same group can be understood as $\mathcal{T}$ dichromatic or as $\mathcal{Z}$ dichromatic, though based on different monochromatic groups $G$ and $\tilde G$, respectively. However, the $\mathcal{Z}$ dichromatic groups in Tables \ref{table:G.summary} and \ref{table:G.dichro} are nevertheless distinct from the $\mathcal{T}$ dichromatic groups since the groups $\tilde G$ are not one of the structural point groups in all these cases.

Further groups can be constructed by taking a group that is dichromatic with respect to $\mathcal{T}$ or $\mathcal{Z}$ or pseudo-dichromatic with respect to $\Omega$ and add one generator. If it is the same element we do not obtain new groups:
\begin{align}
M &= H + (G\setminus H)A + [H + (G\setminus H)A]A \nonumber \\
&= H + G\setminus H + HA + (G\setminus H)A \nonumber \\
&= G + GA ,
\end{align}
where $A \in \{\mathcal{T}, \mathcal{Z}, \Omega\}$. If it is a different generator we instead find
\begin{align}
M &= H + (G\setminus H)A + [H + (G\setminus H)A]B \nonumber \\
&= H + HB + (G\setminus H)C + (G\setminus H)BC \nonumber \\
&= H + (G\setminus H)C + [H + (G\setminus H)C]B ,
\end{align}
where $A, B \in \{\mathcal{T}, \mathcal{Z}, \Omega\}$ and $AB = C \in \{\mathcal{T}, \mathcal{Z}, \Omega\}$. We see that $A$ and $C$ can be interchanged without generating a new group. Furthermore, choosing $B$ fixes the $A$ and $C$ up to their irrelevant order since $A$, $B$, and $C$ are distinct by construction. Hence, for each structural group $G$ and subgroup $H$, there are three new groups distinguished by the choice of $B$. These groups can be written as (the two forms corresponding to the interchange of $A$ and $C$ are both given in all three cases)
\begin{align}
&H + (G\setminus H)\mathcal{Z} + [H + (G\setminus H)\mathcal{Z}]\mathcal{T} \nonumber \\
&\quad= H + (G\setminus H)\Omega + [H + (G\setminus H)\Omega]\mathcal{T} ,
\label{Agroups.mixed1} \\
&H + (G\setminus H)\mathcal{T} + [H + (G\setminus H)\mathcal{T}]\mathcal{Z} \nonumber \\
&\quad= H + (G\setminus H)\Omega + [H + (G\setminus H)\Omega]\mathcal{Z} ,
\label{Agroups.mixed2} \\
&H + (G\setminus H)\mathcal{T} + [H + (G\setminus H)\mathcal{T}]\Omega \nonumber \\
&\quad= H + (G\setminus H)\mathcal{Z} + [H + (G\setminus H)\mathcal{Z}]\Omega \nonumber \\
&\quad= H + H\Omega + [G\setminus H + (G\setminus H)\Omega]\mathcal{T} \nonumber \\
&\quad= H + H\Omega + [G\setminus H + (G\setminus H)\Omega]\mathcal{Z} .
\label{Aground.mixed3}
\end{align}
The first class is obviously $\mathcal{T}$ gray, the second is $\mathcal{Z}$ gray, and the third is both. These $33$ new groups are listed in Table~\ref{table:G.mixed}. If we instead start from a dichromatic or pseudo-dichromatic group and add any two of $\mathcal{T}$, $\mathcal{Z}$, or $\Omega$, the third element is also an element of the group. This does not produce a new group since
\begin{align}
M &= H + (G\setminus H)A + [H + (G\setminus H)A]\mathcal{T} \nonumber \\
&\quad{}+ [H + (G\setminus H)A]\mathcal{Z}
  + [H + (G\setminus H)A]\Omega \nonumber \\
&= G + G\mathcal{T} + G\mathcal{Z} + G\Omega ,
\end{align}
where $A \in \{\mathcal{T}, \mathcal{Z}, \Omega\}$.

\begin{table}
    \caption{\label{table:G.mixed}Mixed (pseudo-)dichromatic and gray CBED groups. See Eqs.\ (\ref{Agroups.mixed1})--(\ref{Aground.mixed3}) for the construction scheme based on the structural groups.}
    \begin{ruledtabular}
    \begin{tabular}{ccc}
    $\mathcal{T}$ gray groups & $\mathcal{Z}$ gray groups &
    $\mathcal{T}$ and $\mathcal{Z}$ gray groups \\ \hline
    $2_R1'$ & $2'1_R$ & $2'1'_R$ \\
    $4_R1'$ & $4'1_R$ & $4'1'_R$ \\
    $6_R1'$ & $6'1_R$ & $6'1'_R$ \\
    $m_R1'$ & $m'1_R$ & $m'1'_R$ \\
    $2_Rmm_R1'$ & $2'mm'1_R$ & $2'mm'1'_R$ \\
    $2m_Rm_R1'$ & $2m'm'1_R$ & $2m'm'1'_R$ \\
    $3m_R1'$ & $3m'1_R$ & $3m'1'_R$ \\
    $4_Rmm_R1'$ & $4'mm'1_R$ & $4'mm'1'_R$ \\
    $4m_Rm_R1'$ & $4m'm'1_R$ & $4m'm'1'_R$ \\
    $6_Rmm_R1'$ & $6'mm'1_R$ & $6'mm'1'_R$ \\
    $6m_Rm_R1'$ & $6m'm'1_R$ & $6m'm'1'_R$
    \end{tabular}
    \end{ruledtabular}
\end{table}

The final class of magnetic CBED groups emerges if one multiplies the coset $G\setminus H_1$ for a halving subgroup $H_1$ by $A \in \{\mathcal{T}, \mathcal{Z}, \Omega\}$ and the coset $G\setminus H_2$ for another halving subgroup $H_2$ by $B \in \{\mathcal{T}, \mathcal{Z}, \Omega\}$. If $H_1$ and $H_2$ are equal one does not obtain new groups because each of the operations $\mathcal{T}$, $\mathcal{Z}$, and $\Omega$ squares to the identity and the product of any two of them is the third.

Hence, we require the structural point group $G$ to have two distinct halving subgroups. This applies to three of the relevant groups: $D_2 = 2mm$, $D_4 = 4mm$, and $D_6 = 6mm$, which each have exactly two distinct halving subgroups. In all three cases, the intersection of the two subgroups $H_1$ and $H_2$ is a halving subgroup of both $H_1$ and $H_2$ and is thus a quartering subgroup of $G$. We denote this subgroup by $Q = H_1 \cap H_2$. The only quartering subgroups that appear are $C_1$, $C_2$, and $C_3$.

Let $h_1$ be a generator of $H_1$ that is not an element of $H_2$ and let $h_2$ be a generator of $H_2$ that is not an element of $H_1$. Hence, $h_1$ and $h_2$ are not elements of $Q$. Then $G$ can be decomposed into cosets according to
\begin{align}
G &= Q + Q h_1 + Q h_2 + Q h_1h_2 \nonumber \\
&= Q + Q h_1 + Q h_2 + Q h_2h_1 .
\end{align}
The last equality also holds if $h_1$ and $h_2$ do not commute since $Q h_2h_1$ is just a reordering of $Q h_1h_2$. Moreover, we find
\begin{align}
Q h_1 + Q h_2h_1 &= H_2 h_1 = G \setminus H_2 , \\
Q h_2 + Q h_1h_2 &= H_1 h_2 = G \setminus H_1 .
\end{align}
By multiplying these complements by \emph{the same} element $A \in \{\mathcal{T}, \mathcal{Z}, \Omega\}$, we do not obtain new groups since
\begin{align}
M &= Q + Q h_1 A + Q h_2 A + Q h_1 h_2 A^2 \nonumber \\
&= Q + Q h_1 h_2 + (Q h_1 + Q h_2)A \nonumber \\
&= Q + Q h_1 h_2 + [G \setminus (Q + Q h_1 h_2)]A
\end{align}
is a dichromatic or pseudo-dichromatic group already contained in Table~\ref{table:G.dichro}.

By instead multiplying the complements by distinct elements $A, B \in \{\mathcal{T}, \mathcal{Z}, \Omega\}$, we obtain the groups
\begin{equation}
M = Q + Q h_1 A + Q h_2 B + Q h_1 h_2 C ,
\end{equation}
where $C = AB \in \{\mathcal{T}, \mathcal{Z}, \Omega\}$. Interchanging $h_1$ and $h_2$ has the same effect as interchanging $A$ and $B$. Here, we fix $h_1$ and $h_2$ and consider all choices for $A$ and $B$. Specifically, we choose $h_1$ to be the generator $c_n$, $n=2,4,6$, of rotations (we use lower-case ``\textit{c}'' to avoid confusion with the groups $C_n$) and $h_2$ as the mirror reflection $m_x$, choosing the \textit{x}-axis normal to the mirror plane. The resulting groups are
\begin{equation}
M = Q + Q c_n A + Q m_x B + Q c_n m_x C ,
\end{equation}
where $Q$ is one of the groups $C_1$, $C_2$, and $C_3$. The explicit groups $M$ are
\begin{align}
&\mbox{from $D_2$:} && \big\{ 1, c_2A, m_xB, m_yC \big\} , \\
&\mbox{from $D_4$:} && \big\{ 1, c_4A, c_2, c_4^{-1}A, \nonumber \\
&&& m_xB, c_4m_xC, c_2m_xB, c_4^{-1}m_xC \big\} , \\
&\mbox{from $D_6$:} && \big\{ 1, c_6A, c_3, c_2A, c_3^{-1}, c_6^{-1}A, m_xB, c_6m_xC,
  \nonumber \\
&&& c_3m_xB, c_2m_xC, c_3^{-1}m_xB, c_6^{-1}m_xC \big\} .
\end{align}
We see that by a suitable rotation about the \textit{z}-axis, $B$ and $C$ are interchanged. This rotation is not considered to lead to a distinct group. The three choices for $A$ together with the three choices of the structural group $D_2$, $D_4$, and $D_6$ (or of the quartering group $C_1$, $C_2$, and $C_3$) generate nine new groups, which are listed in Table \ref{table:G.double}. These groups are dichromatic since half of the elements are antiunitary. We call them ``double dichromatic'' because they are dichromatic with respect to both $\mathcal{T}$ and $\mathcal{Z}$ involving distinct decompositions of the structural group.

\begin{table}
    \caption{\label{table:G.double}Double dichromatic CBED groups. The construction scheme is discussed in the text.}
    \begin{ruledtabular}
    \begin{tabular}{ccc}
    \multicolumn{3}{c}{Double dichromatic groups} \\ \hline
    $2'm'_Rm_R$ & $2_Rm'_Rm'$ & $2'_Rm'm_R$ \\
    $4'm'_Rm_R$ & $4_Rm'_Rm'$ & $4'_Rm'm_R$ \\
    $6'm'_Rm_R$ & $6_Rm'_Rm'$ & $6'_Rm'm_R$
    \end{tabular}
    \end{ruledtabular}
\end{table}

\section{Mappings of magnetic point groups to magnetic CBED groups} \label{app:mPGvsmDG}

In this Appendix, we list the mapping from the 122 three-dimensional magnetic point groups to the 125 magnetic CBED groups for all distinct beam directions. The tables are organized according to crystal classes: The mappings for triclinic point groups are shown in Table \ref{table:Triclinic}, those for monoclinic groups in Table \ref{table:Monoclinic}, those for orthorhomic groups in Table \ref{table:Orthorhombic}, those for tetragonal groups in Tables \ref{table:Tetragonal1} and \ref{table:Tetragonal2}, those for trigonal groups in Tables \ref{table:Trigonal1} and \ref{table:Trigonal2}, those for hexagonal groups in Tables \ref{table:Hexagonal1} and \ref{table:Hexagonal2}, and those for cubic groups in Tables \ref{table:Cubic1} and \ref{table:Cubic2}. In all tables, the mapping of magnetic point groups (mPG) to magnetic CBED groups (magnetic diffraction groups, mDG) is shown for all slab orientations. The underlying structural point group is given in Hermann-Mauguin and, at the first occurrence, also in Schoenflies notation. The slab normal is parallel to the beam direction, which is denoted by $[uvw]$ in crystallographic coordinates. Here, variables $u$, $v$, $w$ stand for generic values, which are distinct from the special values given in the tables to denote high-symmetry directions. All tables also show the corresponding structural point group (PG). We note that there are occasional permutations of the symmetry symbols in the point group notations, e.g., $mm2$ instead of $2mm$, corresponding to different choices of unit cell orientations. By using them we follow established conventions in the literature \cite{Buxton1976, Tanaka2010}. These permutations do not affect the symmetry considerations.

\begin{table}
\centering
 \caption{\label{table:Triclinic}Assignment of triclinic magnetic point groups to magnetic CBED groups.}
 \begin{ruledtabular}
    \begin{tabular}{ccc}
		PG & mPG & \multicolumn{1}{c}{mDG for beam directions}\\ 
		\multicolumn{2}{c}{ }  & $[uvw]$ \\ 
		\hline
		$1$ ($C_1$) & $1$  & $1$ \\   
		$1$ & $11'$ & $11'$ \\  
		$\bar{1}$ ($C_i$) & $\bar{1}$ & $2_R$\\  
		$\bar{1}$ & $\bar{1}1'$ & $2_R1'$\\  
		$\bar{1}$ & $\bar{1}'$ & $2'_R$\\  
		
	\end{tabular}
 \end{ruledtabular}
\end{table}


\begin{table}
    \caption{\label{table:Monoclinic}Assignment of monoclinic magnetic point groups to magnetic CBED groups.}
    \begin{ruledtabular}
    \begin{tabular}{ccccc}
		PG & mPG & \multicolumn{3}{c}{mDG for beam directions}\\ 
		 \multicolumn{2}{c}{ }  & $[010]$ & $[u0w]$ & $[uvw]$\\ 
		\hline
		$2$ ($C_2$) & $2$  & $2$ & $m_R$ & $1$\\   	
		$2$ & $21'$ & $21'$ & $m_R1'$ & $11'$\\  
		$2$ & $2'$ & $2'$ & $m'_R$ & $1$\\  
		$m$ ($C_s$) & $m$ & $1_R$ & $m$ & $1$\\  
		$m$ & $m1'$ & $1_R1'$ & $m1'$ & $11'$\\  
		$m$ & $m'$ & $1'_R$ & $m'$ & $1$\\  
		$2/m$ ($C_{2h}$) & $2/m$ & $21_R$ & $2_Rmm_R$ & $2_R$\\  
        $2/m$ & $2/m 1'$ & $21_R1'$ & $2_Rmm_R1'$ & $2_R1'$\\  
		$2/m$ & $2/m'$ & $21'_R$ & $2'_Rm'm_R$ & $2'_R$\\  
		$2/m$ & $2'/m'$ & $2'1'_R$ & $2_Rm'm'_R$ & $2_R$\\  
		$2/m$ & $2'/m$ & $2'1_R$ & $2'_Rmm'_R$ & $2'_R$\\  
	\end{tabular}
    \end{ruledtabular}
\end{table}


\begin{table*}
    \caption{\label{table:Orthorhombic}Assignment of orthorhombic magnetic point groups to magnetic CBED groups.}
    \begin{ruledtabular}
    \begin{tabular}{ccccccccc}
		PG & mPG & \multicolumn{7}{c}{mDG for beam directions}\\ 
		 \multicolumn{2}{c}{ }  & $[010]$ & $[001]$ & $[100]$ & $[u0w]$ & $[uv0]$ & $[0vw]$ & $[uvw]$\\ 
		\hline
		$222$ ($D_{2}$) & $222$  & $2m_Rm_R$ & $2m_Rm_R$ & $2m_Rm_R$  & $m_R$ & $m_R$ & $m_R$ & $1$\\   
		$222$ & $2221'$  & $2m_Rm_R1'$ & $2m_Rm_R1'$ & $2m_Rm_R1'$  & $m_R1'$ & $m_R1'$ & $m_R1'$ & $11'$\\    
        $222$ & $2'2'2$  & $2'm_Rm'_R$ & $2m'_Rm'_R$ & $2'm_Rm'_R$  & $m'_R$ & $m_R$ & $m'_R$ & $1$\\ 
         $mm2$ ($C_{2v}$) & $mm2$  & $m1_R$ & $2mm$ & $m1_R$  & $m$ & $m_R$ & $m$ & $1$\\   
         $mm2$ & $mm21'$  & $m1_R1'$ & $2mm1'$ & $m1_R1'$  & $m1'$ & $m_R1'$ & $m1'$ & $11'$\\   
        $mm2$ & $m'm'2$  & $m'1'_R$ & $2m'm'$ & $m'1'_R$  & $m'$ & $m_R$ & $m'$ & $1$\\   
        $mm2$ & $m'm2'$  & $m'1_R$ & $2'mm'$ & $m1'_R$  & $m$ & $m'_R$ & $m'$ & $1$\\  
        $mmm$ ($D_{2h}$) & $mmm$  & $2mm1_R$ & $2mm1_R$ & $2mm1_R$  & $2_Rmm_R$ & $2_Rmm_R$ & $2_Rmm_R$ & $2_R$\\  
        $mmm$ & $mmm1'$  & $2mm1_R1'$ & $2mm1_R1'$ & $2mm1_R1'$  & $2_Rmm_R1'$ & $2_Rmm_R1'$ & $2_Rmm_R1'$ & $2_R1'$\\   
        $mmm$ & $m'm'm$  & $2'mm'1'_R$ & $2m'm'1_R$ & $2'mm'1'_R$  & $2_Rm'm'_R$ & $2_Rmm_R$ & $2_Rm'm'_R$ & $2_R$\\    
        $mmm$ & $m'm'm'$  & $2m'm'1'_R$ & $2m'm'1'_R$ & $2m'm'1'_R$  & $2'_Rm'm_R$ & $2'_Rm'm_R$ & $2'_Rm'm_R$ & $2'_R$\\   
        $mmm$ & $mmm'$  & $2'mm'1_R$ & $2mm1'_R$ & $2'mm'1_R$  & $2'_Rmm'_R$ & $2'_Rm'm_R$ & $2'_Rmm'_R$ & $2'_R$\\    
	\end{tabular}
    \end{ruledtabular}
\end{table*}


\begin{table}
    \caption{\label{table:Tetragonal1}Assignment of tetragonal magnetic point groups to magnetic CBED groups.}
    \begin{ruledtabular}
    \begin{tabular}{ccccc}
		PG & mPG & \multicolumn{3}{c}{mDG for beam directions}\\ 
		\multicolumn{2}{c}{ }  & $[001]$ & $[uv0]$ & $[uvw]$ \\ 
		\hline
		$4$ ($C_4$) & $4$ & $4$ & $m_R$ & $1$\\   
        $4$ & $41'$ & $41'$ & $m_R1'$ & $11'$\\   
        $4$ & $4'$ & $4'$ & $m_R$ & $1$\\ 
        $\bar{4}$ ($S_4$) & $\bar{4}$ & $4_R$ & $m_R$ & $1$\\ 
        $\bar{4}$ & $\bar{4}1'$ & $4_R1'$ & $m_R1'$ & $11'$\\ 
        $\bar{4}$ & $\bar{4}'$ & $4'_R$ & $m_R$ & $1$\\ 
        $4/m$ ($C_{4h}$) & $4/m$ & $41_R$ & $2_Rmm_R$ & $2_R$\\ 
        $4/m$ & $4/m1'$ & $41_R1'$ & $2_Rmm_R1'$ & $2_R1'$\\ 
        $4/m$ & $4'/m$ & $4'1_R$ & $2_Rmm_R$ & $2_R$\\  
        $4/m$ & $4/m'$ & $41'_R$ & $2'_Rm'm_R$ & $2'_R$\\ 
        $4/m$ & $4'/m'$ & $4'1'_R$ & $2'_Rm'm_R$ & $2'_R$\\  
	\end{tabular}
    \end{ruledtabular}
\end{table}

\begin{table*}
    \caption{\label{table:Tetragonal2}Assignment of tetragonal magnetic point groups to magnetic CBED groups (continued).}
    \begin{ruledtabular}
    \begin{tabular}{ccccccccc}
		PG & mPG & \multicolumn{7}{c}{mDG for beam directions}\\ 
		 \multicolumn{2}{c}{ }  & $[001]$ & $\langle100\rangle$ & $\langle110\rangle$ & $[u0w]$ & $[uv0]$ & $[uuw]$ & $[uvw]$\\ 
		\hline
		$422$ ($D_4$) & $422$ & $4m_Rm_R$ & $2m_Rm_R$ & $2m_Rm_R$ & $m_R$ & $m_R$ & $m_R$ & $1$\\   
  	$422$ & $4221'$ & $4m_Rm_R1'$ & $2m_Rm_R1'$ & $2m_Rm_R1'$ & $m_R1'$ & $m_R1'$ & $m_R1'$ & $11'$\\   
  	$422$ & $4'22'$ & $4'm_Rm'_R$ & $2m_Rm_R$ & $2'm_Rm'_R$ & $m_R$ & $m_R$ & $m'_R$ & $1$\\ 
  	$422$ & $42'2'$ & $4m'_Rm'_R$ & $2'm_Rm'_R$ & $2'm_Rm'_R$ & $m_R'$ & $m_R$ & $m'_R$ & $1$\\ 
  	$4mm$ ($C_{4v}$) & $4mm$ & $4mm$ & $m1_R$ & $m1_R$ & $m$ & $m_R$ & $m$ & $1$\\   
  	$4mm$ & $4mm1'$ & $4mm1'$ & $m1_R1'$ & $m1_R1'$ & $m1'$ & $m_R1'$ & $m1'$ & $11'$\\    
  	$4mm$ & $4'mm'$ & $4'mm'$ & $m1_R$ & $m'1'_R$ & $m$ & $m_R$ & $m'$ & $1$\\  
  	$4mm$ & $4m'm'$ & $4m'm'$ & $m'1'_R$ & $m'1'_R$ & $m'$ & $m_R$ & $m'$ & $1$\\   
  	$\bar{4}2m$ ($D_{2d}$) & $\bar{4}2m$ & $4_Rmm_R$ & $2m_Rm_R$ & $m1_R$ & $m_R$ & $m_R$ & $m$ & $1$\\ 
  	$\bar{4}2m$ & $\bar{4}2m1'$ & $4_Rmm_R1'$ & $2m_Rm_R1'$ & $m1_R1'$ & $m_R1'$ & $m_R1'$ & $m1'$ & $11'$\\   
  	$\bar{4}2m$ & $\bar{4}'2m'$ & $4'_Rm'm_R$ & $2m_Rm_R$ & $m'1'_R$ & $m_R$ & $m_R$ & $m'$ & $1$\\   
  	$\bar{4}2m$ & $\bar{4}'2'm$ & $4'_Rmm'_R$ & $2'm_Rm'_R$ & $m1_R$ & $m'_R$ & $m_R$ & $m$ & $1$\\   
  	$\bar{4}2m$ & $\bar{4}2'm'$ & $4_Rm'm'_R$ & $2'm_Rm'_R$ & $m'1'_R$ & $m'_R$ & $m_R$ & $m'$ & $1$\\  
  	$4/mmm$ ($D_{4h}$) & $4/mmm$ & $4mm1_R$ & $2mm1_R$ & $2mm1_R$ & $2_Rmm_R$ & $2_Rmm_R$ & $2_Rmm_R$ & $2_R$\\   
  	$4/mmm$ & $4/mmm1'$ & $4mm1_R1'$ & $2mm1_R1'$ & $2mm1_R1'$ & $2_Rmm_R1'$ & $2_Rmm_R1'$ & $2_Rmm_R1'$ & $2_R1'$\\ 
  	$4/mmm$ & $4'/mmm'$ & $4'mm'1_R$ & $2mm1_R$ & $2'mm'1'_R$ & $2_Rmm_R$ & $2_Rmm_R$ & $2_Rm'm'_R$ & $2_R$\\  
  	$4/mmm$ & $4/mm'm'$ & $4m'm'1_R$ & $2'mm'1'_R$ & $2'mm'1'_R$ & $2_Rm'm'_R$ & $2_Rmm_R$ & $2_Rm'm'_R$ & $2_R$\\ 
  	$4/mmm$ & $4/m'm'm'$ & $4m'm'1'_R$ & $2m'm'1'_R$ & $2m'm'1'_R$ & $2'_Rm'm_R$ & $2'_Rm'm_R$ & $2'_Rm'm_R$ & $2'_R$\\   
  	$4/mmm$ & $4/m'mm$ & $4mm1'_R$ & $2'mm'1_R$ & $2'mm'1_R$ & $2'_Rmm'_R$ & $2'_Rm'm_R$ & $2'_Rmm'_R$ & $2'_R$\\   
  	$4/mmm$ & $4'/m'm'm$ & $4'mm'1'_R$ & $2m'm'1'_R$ & $2'mm'1_R$ & $2'_Rm'm_R$ & $2'_Rm'm_R$ & $2'_Rmm'_R$ & $2'_R$\\    
	\end{tabular}
    \end{ruledtabular}
\end{table*}


\begin{table}
    \caption{\label{table:Trigonal1}Assignment of trigonal magnetic point groups to magnetic CBED groups.}
    \begin{ruledtabular}
    \begin{tabular}{cccc}
		PG & mPG & \multicolumn{2}{c}{mDG for beam directions}\\ 
		 \multicolumn{2}{c}{ }  & $[0001]$ & $[uv.w]$ \\ 
		\hline
		$3$ ($C_3$) & $3$  & $3$ & $1$ \\ 
		$3$ & $31'$  & $31'$ & $11'$ \\  
        $\bar{3}$ ($C_{3i}$) & $\bar{3}$  & $6_R$ & $2_R$ \\   
        $\bar{3}$ & $\bar{3}1'$  & $6_R1'$ & $2_R1'$ \\   
        $\bar{3}$ & $\bar{3}'$  & $6'_R$ & $2'_R$ \\   
	\end{tabular}
    \end{ruledtabular}
\end{table}

\begin{table}
    \caption{\label{table:Trigonal2}Assignment of trigonal magnetic point groups to magnetic CBED groups (continued).}
    \begin{ruledtabular}
    \begin{tabular}{cccccc}
		PG & mPG & \multicolumn{4}{c}{mDG for beam directions}\\ 
		 \multicolumn{2}{c}{ } & $[0001]$ & $\langle11\bar{2}0\rangle$ & $[u\bar{u}.w]$ & $[uv.w]$\\ 
		\hline
		$32$ ($D_3$) & $32$ & $3m_R$ & $2$ & $m_R$ & $1$\\   
  	$32$ & $321'$ & $3m_R1'$ & $21'$ & $m_R1'$ & $11'$\\   
  	$32$ & $32'$ & $3m'_R$ & $2'$ & $m'_R$ & $1$\\   
  	$3m$ ($C_{3v}$) & $3m$ & $3m$ & $1_R$ & $m$ & $1$\\   
  	$3m$ & $3m1'$ & $3m1'$ & $1_R1'$ & $m1'$ & $11'$\\   
  	$3m$ & $3m'$ & $3m'$ & $1'_R$ & $m'$ & $1$\\
  	$\bar{3}m$ ($D_{3d}$) & $\bar{3}m$ & $6_Rmm_R$ & $21_R$ & $2_Rmm_R$ & $2_R$\\   
  	$\bar{3}m$ & $\bar{3}m1'$ & $6_Rmm_R1'$ & $21_R1'$ & $2_Rmm_R1'$ & $2_R1'$\\   
  	$\bar{3}m$ & $\bar{3}m'$ & $6_Rm'm'_R$ & $2'1'_R$ & $2_Rm'm'_R$ & $2_R$\\   
  	$\bar{3}m$ & $\bar{3}'m'$ & $6'_Rm'm_R$ & $21'_R$ & $2'_Rm'm_R$ & $2'_R$\\   
  	$\bar{3}m$ & $\bar{3}'m$ & $6'_Rmm'_R$ & $2'1_R$ & $2'_Rmm'_R$ & $2'_R$\\   
	\end{tabular}
    \end{ruledtabular}
\end{table}

\begin{table}
    \caption{\label{table:Hexagonal1}Assignment of hexagonal magnetic point groups to magnetic CBED groups.}
    \begin{ruledtabular}
    \begin{tabular}{ccccc}
		PG & mPG & \multicolumn{3}{c}{mDG for beam directions}\\ 
		 \multicolumn{2}{c}{ }  & $[0001]$ & $[uv.0]$ & $[uv.w]$ \\ 
		\hline
		$6$ ($C_6$) & $6$  & $6$ & $m_R$ & $1$ \\   
  	$6$ & $61'$  & $61'$ & $m_R1'$ & $11'$ \\   
  	$6$ & $6'$  & $6'$ & $m'_R$ & $1$ \\
  	$\bar{6}$ ($C_{3h}$) & $\bar{6}$  & $31_R$ & $m$ & $1$ \\   
  	$\bar{6}$ & $\bar{6}1'$  & $31_R1'$ & $m1'$ & $11'$ \\   
  	$\bar{6}$ & $\bar{6}'$  & $31'_R$ & $m'$ & $1$ \\   
  	$6/m$ ($C_{6h}$) & $6/m$  & $61_R$ & $2_Rmm_R$ & $2_R$ \\   
  	$6/m$ & $6/m1'$  & $61_R1'$ & $2_Rmm_R1'$ & $2_R1'$ \\   
  	$6/m$ & $6'/m'$  & $6'1'_R$ & $2_Rm'm'_R$ & $2_R$ \\   
  	$6/m$ & $6/m'$  & $61'_R$ & $2'_Rm'm_R$ & $2'_R$ \\   
  	$6/m$ & $6'/m$  & $6'1_R$ & $2'_Rmm'_R$ & $2'_R$ \\   
	\end{tabular}
    \end{ruledtabular}
\end{table}

\begin{table*}
    \caption{\label{table:Hexagonal2}Assignment of hexagonal magnetic point groups to magnetic CBED groups (continued).}
    \begin{ruledtabular}
    \begin{tabular}{ccccccccc}
		PG & mPG & \multicolumn{7}{c}{mDG for beam directions}\\ 
		 \multicolumn{2}{c}{ } & $[0001]$ & $\langle11\bar{2}0\rangle$ & $\langle1\bar{1}00\rangle$ & $[uv.0]$ & $[uu.w]$ & $[u\bar{u}.w]$ & $[uv.w]$\\ 
		\hline
		$622$ ($D_6$) & $622$ & $6m_Rm_R$ & $2m_Rm_R$ & $2m_Rm_R$ & $m_R$ & $m_R$ & $m_R$ & $1$\\   
  	$622$ & $6221'$ & $6m_Rm_R1'$ & $2m_Rm_R1'$ & $2m_Rm_R1'$ & $m_R1'$ & $m_R1'$ & $m_R1'$ & $11'$\\    
  	$622$ & $6'22'$ & $6'm_Rm'_R$ & $2m'_Rm'_R$ & $2'm_Rm'_R$ & $m'_R$ & $m'_R$ & $m_R$ & $1$\\ 
        $622$ & $62'2'$ & $6m'_Rm'_R$ & $2'm_Rm'_R$ & $2'm_Rm'_R$ & $m_R$ & $m'_R$ & $m'_R$ & $1$\\
  	$6mm$ ($C_{6v}$) & $6mm$ & $6mm$ & $m1_R$ & $m1_R$ & $m_R$ & $m$ & $m$ & $1$\\   
  	$6mm$ & $6mm1'$ & $6mm1'$ & $m1_R1'$ & $m1_R1'$ & $m_R1'$ & $m1'$ & $m1'$ & $11'$\\   
  	$6mm$ & $6'mm'$ & $6'mm'$ & $m'1_R$ & $m1'_R$ & $m'_R$ & $m'$ & $m$ & $1$\\   
  	$6mm$ & $6m'm'$ & $6m'm'$ & $m'1'_R$ & $m'1'_R$ & $m_R$ & $m'$ & $m'$ & $1$\\   
  	$\bar{6}m2$ ($D_{3h}$) & $\bar{6}m2$ & $3m1_R$ & $m1_R$ & $2mm$ & $m$ & $m_R$ & $m$ & $1$\\   
  	$\bar{6}m2$ & $\bar{6}m21'$ & $3m1_R1'$ & $m1_R1'$ & $2mm1'$ & $m1'$ & $m_R1'$ & $m1'$ & $11'$\\   
  	$\bar{6}m2$ & $\bar{6}'m'2$ & $3m'1'_R$ & $m'1'_R$ & $2m'm'$ & $m'$ & $m_R$ & $m'$ & $1$\\   
  	$\bar{6}m2$ & $\bar{6}'m2'$ & $3m1'_R$ & $m'1_R$ & $2'mm'$ & $m'$ & $m'_R$ & $m$ & $1$\\ 
  	$\bar{6}m2$ & $\bar{6}m'2'$ & $3m'1_R$ & $m1'_R$ & $2'mm'$ & $m$ & $m'_R$ & $m'$ & $1$\\ 
  	$6/mmm$ ($D_{6h}$) & $6/mmm$ & $6mm1_R$ & $2mm1_R$ & $2mm1_R$ & $2_Rmm_R$ & $2_Rmm_R$ & $2_Rmm_R$ & $2_R$\\ 
  	$6/mmm$ & $6/mmm1'$ & $6mm1_R1'$ & $2mm1_R1'$ & $2mm1_R1'$ & $2_Rmm_R1'$ & $2_Rmm_R1'$ & $2_Rmm_R1'$ & $2_R1'$\\   
  	$6/mmm$ & $6'/m'mm'$ & $6'mm'1'_R$ & $2m'm'1_R$ & $2'mm'1'_R$ & $2_Rm'm'_R$ & $2_Rm'm'_R$ & $2_Rmm_R$ & $2_R$\\   
  	$6/mmm$ & $6/mm'm'$ & $6m'm'1_R$ & $2'mm'1'_R$ & $2'mm'1'_R$ & $2_Rmm_R$ & $2_Rm'm'_R$ & $2_Rm'm'_R$ & $2_R$\\   
  	$6/mmm$ & $6/m'm'm'$ & $6m'm'1'_R$ & $2m'm'1'_R$ & $2m'm'1'_R$ & $2'_Rm'm_R$ & $2'_Rm'm_R$ & $2'_Rm'm_R$ & $2'_R$\\   
  	$6/mmm$ & $6/m'mm$ & $6mm1'_R$ & $2'mm'1_R$ & $2'mm'1_R$ & $2'_Rm'm_R$ & $2'_Rmm'_R$ & $2'_Rmm'_R$ & $2'_R$\\   
  	$6/mmm$ & $6'/mmm'$ & $6'mm'1_R$ & $2'mm'1_R$ & $2mm1'_R$ & $2'_Rmm'_R$ & $2'_Rm'm_R$ & $2'_Rmm'_R$ & $2'_R$\\   
	\end{tabular}
    \end{ruledtabular}
\end{table*}


\begin{table}
    \caption{\label{table:Cubic1}Assignment of cubic magnetic point groups to magnetic CBED groups.}
    \begin{ruledtabular}
    \begin{tabular}{cccccc}
		PG & mPG & \multicolumn{4}{c}{mDG for beam directions}\\ 
		 \multicolumn{2}{c}{ } & $\langle100\rangle$ & $\langle111\rangle$ & $\langle uv0\rangle$ & $[uvw]$ \\ 
		\hline
		$23$ ($T$) & $23$ & $2m_Rm_R$ & $3$ & $m_R$ & $1$\\   
  	$23$ & $231'$ & $2m_Rm_R1'$ & $31'$ & $m_R1'$ & $11'$\\   
  	$m\bar{3}$ ($T_h$) & $m\bar{3}$ & $2mm1_R$ & $6_R$ & $2_Rmm_R$ & $2_R$\\   
  	$m\bar{3}$ & $m\bar{3}1'$ & $2mm1_R1'$ & $6_R1'$ & $2_Rmm_R1'$ & $2_R1'$\\   
  	$m\bar{3}$ & $m'\bar{3}'$ & $2m'm'1'_R$ & $6'_R$ & $2'_Rm'm_R$ & $2'_R$\\   
	\end{tabular}
    \end{ruledtabular}
\end{table}

\begin{table*}
    \caption{\label{table:Cubic2}Assignment of cubic magnetic point groups to magnetic CBED groups (continued).}
    \begin{ruledtabular}
    \begin{tabular}{cccccccc}
		PG & mPG & \multicolumn{6}{c}{mDG for beam directions}\\ 
		 \multicolumn{2}{c}{ }   & $\langle100\rangle$ & $\langle110\rangle$ & $\langle111\rangle$ & $\langle uv0\rangle$ & $\langle uuw\rangle$ & $[uvw]$\\ 
		\hline
        $432$ ($O$) & $432$ & $4m_Rm_R$ & $2m_Rm_R$ & $3m_R$ & $m_R$ & $m_R$ & $1$\\
		$432$ & $4321'$ & $4m_Rm_R1'$ & $2m_Rm_R1'$ & $3m_R1'$ & $m_R1'$ & $m_R1'$ & $11'$\\   
  	$432$ & $4'32'$ & $4'm_Rm'_R$ & $2'm_Rm'_R$ & $3m'_R$ & $m_R$ & $m'_R$ & $1$\\   
  	$\bar{4}3m$ ($T_d$) & $\bar{4}3m$ & $4_Rmm_R$ & $m1_R$ & $3m$ & $m_R$ & $m$ & $1$\\   
  	$\bar{4}3m$ & $\bar{4}3m1'$ & $4_Rmm_R1'$ & $m1_R1'$ & $3m1'$ & $m_R1'$ & $m1'$ & $11'$\\   
  	$\bar{4}3m$ & $\bar{4}'3m'$ & $4'_Rm'm_R$ & $m'1'_R$ & $3m'$ & $m_R$ & $m'$ & $1$\\
  	$m\bar{3}m$ ($O_h$) & $m\bar{3}m$ & $4mm1_R$ & $2mm1_R$ & $6_Rmm_R$ & $2_Rmm_R$ & $2_Rmm_R$ & $2_R$\\   
  	$m\bar{3}m$ & $m\bar{3}m1'$ & $4mm1_R1'$ & $2mm1_R1'$ & $6_Rmm_R1'$ & $2_Rmm_R1'$ & $2_Rmm_R1'$ & $2_R1'$\\   
  	$m\bar{3}m$ & $m'\bar{3}'m$ & $4'mm'1'_R$ & $2'mm'1_R$ & $6'_Rmm'_R$ & $2'_Rm'm_R$ & $2'_Rmm'_R$ & $2'_R$\\   
  	$m\bar{3}m$ & $m\bar{3}m'$ & $4'mm'1_R$ & $2'mm'1'_R$ & $6_Rm'm'_R$ & $2_Rmm_R$ & $2_Rm'm'_R$ & $2_R$\\   
  	$m\bar{3}m$ & $m'\bar{3}'m'$ & $4m'm'1'_R$ & $2m'm'1'_R$ & $6'_Rm'm_R$ & $2'_Rm'm_R$ & $2'_Rm'm_R$ & $2'_R$\\  
	\end{tabular}
    \end{ruledtabular}
\end{table*}

\clearpage

\section{Thermal Diffuse Scattering}\label{app:TDS}
\begin{figure}
\centering{\includegraphics{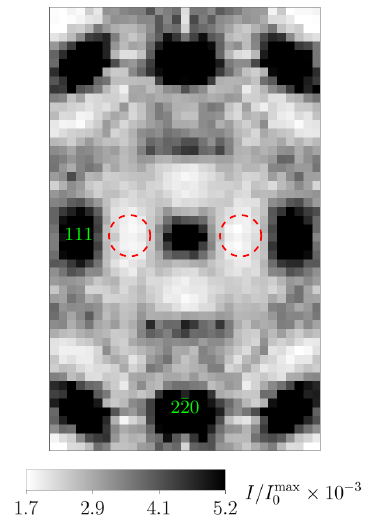}}
\caption{Thermal diffuse background from electron diffraction simulations of nonmagnetic NiO in $\left[11\bar{2}\right]_c$ orientation, visible as intensity at and between strong structural systematic reflections appearing black in the employed gray scale. The red dashed circles indicate the position of the magnetic Bragg reflections. The normalization is identical to the CBED simulations shown in Fig.~\ref{fig:NiO SF}.}
\label{fig:NiO TDS}
\end{figure}

This appendix details the scattering simulations carried out to assess the impact of thermal diffuse scattering (phonon scattering) on magnetic CBED. We have chosen parallel illumination to reduce the computational effort (smaller supercell) considering that the TDS background intensity is similar to the one obtained for converged illumination (CBED), see subsection~\ref{app:NiO}. We performed electron diffraction simulations for NiO including thermal diffuse scattering (TDS) using a frozen-phonon approximation \cite{Loane1991}. Snapshots of the vibrating structure were calculated using molecular dynamics, performed using the atomic simulation environment (\cite{ase-paper}) within Langevin dynamics. The correlated motion of atoms was described using the Orb foundational model as the interatomic potential \cite{ORB}. The assumed orientation of the sample is $\left[11\bar{2}\right]_c$, the acceleration voltage is $300\,\mathrm{kV}$, and the size of the supercell is $29.0292\,\text{\AA} \times 29.6278\,\text{\AA} \times 1000.678\,\text{\AA}$.   The thickness of the sample is the same as in the case of relevant magnetic CBED simulations. The average value of the normalized TDS intensity at the location of magnetic reflections (red dashed circles) is $2.3\times 10^{-3}$, see Fig.~\ref{fig:NiO TDS}.

The comparison with the magnetic scattering simulations (Fig. \ref{fig:NiO SF}), shows that the intensity of the TDS background at the position of the magnetic Bragg peaks is approximately 10 times larger than the magnetic signal. Consequently, in the particular case of purely magnetic reflections in NiO the TDS background determines the acquisition time required to render the measurement of magnetic CBED symmetries significant above the noise level of the background (see Sec. \ref{sec:experiment}).

\section{Beam Tilt}\label{app:Tilt}
\begin{figure}
\centering{\includegraphics{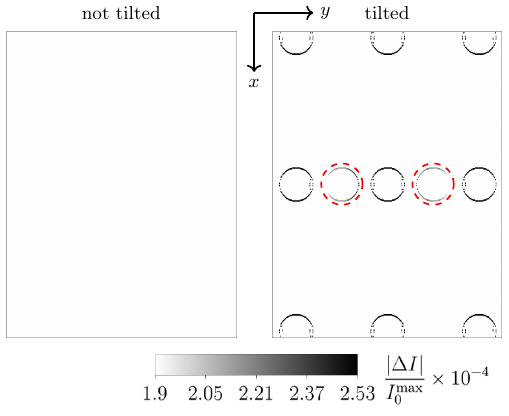}}
\caption{NiO $\left[11\bar{2}\right]_c$ CBED simulations with and without a beam tilt of $67\,\mu\mathrm{rad}$ relative to the $\left[111\right]_c$, i.e., the $y$-axis. Images show differences of intensities of original and $x$-mirror-reflected CBED patterns.}
\label{fig:NiO xtilt}
\end{figure}

This appendix elaborates on the impact of small crystal misorientations, that cannot be avoided experimentally, on the magnetic CBED measurements. To assess its impact, namely to what extent CBED symmetries are affected, we performed CBED simulations of spin flopped NiO in the $\left[11\bar{2}\right]_c$ orientation, including a small beam tilt of $67\,\mu\mathrm{rad}$ relative to the $\left[111\right]_c$ axis, corresponding to the $y$-axis in Fig.~\ref{fig:NiO xtilt}. The figure displays the differences between the original and the $x$-mirror-reflected CBED patterns with tilted illumination in Fig.~\ref{fig:NiO xtilt}(b) and without tilt in Fig.~\ref{fig:NiO xtilt}(a). The tilt results in a slight shift of the CBED discs, visible as very high intensity differences at the edges of original and $x$-reflected CBED discs. Note, however, that the mirror symmetry of the CBED disc's interior is not broken.

\bibliography{mCBED}

\end{document}